\documentclass[reprint,amsmath,amssymb,aps,floatfix,pra]{revtex4-2}

\usepackage{array, makecell}
\usepackage{graphicx}
\usepackage{dcolumn}
\usepackage{bm}
\usepackage{xcolor}
\usepackage{braket}
\usepackage[normalem]{ulem}
\usepackage[utf8]{inputenc}
\usepackage[colorlinks=true,       
            linkcolor=blue,        
            citecolor=blue,        
            urlcolor=blue,         
            backref=true]{hyperref} 

\begin{document}

\title{Nonlinear photonic architecture for fault-tolerant quantum computing}

\author{Maike Ostmann}
\author{Joshua Nunn}
\author{Alex E. Jones}
\email{ajones@orcacomputing.com}
\affiliation{ORCA Computing}

\begin{abstract}
We propose an architecture for fault-tolerant quantum computing that incorporates strong single-photon nonlinearities into a photonic Greenberger-Horne-Zeilinger-measurement-based architecture.
The nonlinearities substantially reduce resource overhead compared to conventional linear-optics-based architectures, which require significant redundancy to accommodate probabilistic photon generation and probabilistic entangling operations. By removing linear-optical failure modes, our nonlinear architecture can also tolerate much higher optical losses than linear approaches, with a baseline loss tolerance of approximately 12\% using a 32-photon resource state and a foliated surface code. Nonlinear photonic architectures provide a route to dramatically improving practical implementations of fault-tolerant quantum computing.
\end{abstract}

\maketitle

\section{Introduction}
Recent years have seen significant progress towards the realisation of fault-tolerant quantum computers, with the advent of more efficient codes \cite{PRXQuantum.2.040101,komoto2024quantum}, and initial implementations of error suppression \cite{Acharya2025}. However, atomic and electronic qubit platforms require high vacuum or cryogenic containment \cite{zhao2025integratedphotonicsplatformhighspeed, PhysRevX.14.041028, CarreraVazquez2024} making it challenging to accommodate the thousands of logical qubits required for large-scale error-corrected algorithms \cite{gidney2025factor2048bitrsa,mohseni2025buildquantumsupercomputerscaling,litinski2023compute256bitellipticcurve}. Modular architectures with electronic qubits networked via photonic or optical interconnects solve this problem \cite{inc2024distributedquantumcomputingsilicon, deGliniasty2024spinopticalquantum, riedel2025scalablephotonicquantuminterconnect}, but these techniques remain immature compared with monolithic demonstrations \cite{weaver2024integrated}.

All-photonic approaches to quantum computing can be natively modular, with fibre networking providing a natural route to large-scale systems. Furthermore, photons do not suffer from environmental thermal, electrical or magnetic noise, and can support high bandwidths. Photonic processors are already being deployed in data centre environments and can be mounted in mostly standard server racks \cite{slysz2025hybrid,AghaeeRad2025,kailasanathan2025quantumenhancedensemblegans}. However, linear-optical architectures rely on probabilistic photodetection to create heralded non-Gaussian entangled states \cite{larsen2025integrated,bartolucci2021creationentangledphotonicstates}. To achieve deterministic operation, each state generator uses nested layers of {multiplexing} \cite{AghaeeRad2025,bartolucci2021switchnetworksphotonicfusionbased}, resulting in very large component counts, and requiring many layers of optical switching, which to date remain too lossy for any demonstrations of error correction. {Alternative platforms based on quantum dots can generate photonic entanglement with substantially reduced multiplexing overhead, but in these schemes the quantum dot itself acts as a single-photon nonlinearity rather than a purely linear optical element.}

We adopt a hybrid approach based on {nonlinear optics}, where strong atom–light coupling mediates photon–photon interactions, enabling the near-deterministic generation of non-Gaussian photonic states. Recent advances at ORCA and other groups \cite{Zektzer:24,bechler2018passive, deGliniasty2024spinopticalquantum, PRXQuantum.5.010102} demonstrate that such strong nonlinearities can support photon–photon coupling and are experimentally achievable. We still employ heralding and a degree of multiplexing, to mitigate the effect of moderate losses. Nevertheless we find that the use of nonlinearities dramatically reduces the number of components and the number of switching layers required. At the same time, we find that standard error correcting codes can accommodate significantly higher optical losses, once the linear-optical resource state generators are replaced with better performing nonlinear generators.

It is not surprising that introducing a photon-photon interaction enables a more efficient approach to computing. Hybrid approaches based on optically networked spin qubits \cite{chan2025practicalblueprint, PRXQuantum.5.010102, inc2024distributedquantumcomputingsilicon, deGliniasty2024spinopticalquantum, dessertaine2026enhancedfaulttolerancephotonicquantum} are essentially leveraging the same physics, combining deterministic entangling operations with photonic interconnects. However, in these approaches, quantum information 'lives' in the spins for many clock cycles, which requires atomic systems with a carefully isolated spin environment. This often entails low-temperature cryogenics with controlled magnetic fields, spin-pumping techniques or high vacuum systems and laser cooling techniques. Optical interfaces with trapped atoms are either lossy \cite{v5k1-whwz, stephenson2020high} or narrowband \cite{krutyanskiy2023entanglement}, and solid-state spin-photon interfaces exhibit large variance in their resonant frequencies over time \cite{deabreu2023waveguide} and between devices \cite{lodahl2015interfacing}. By contrast, an atom-mediated non-linearity acts only when photons impinge on it, and does not require sustained spin coherence. Thus a hybrid system of this kind can retain the advantages of photonics as a platform, i.e. -- modular, fibre-networked, broadband, and environmentally robust -- while sidestepping the challenging hardware overhead and loss requirements of a linear-optical architecture.

In this work, we present a universal fault-tolerant photonic measurement-based quantum computing architecture. Our approach is based on single-photon nonlinearities and $n$-qubit Greenberger-Horne-Zeilinger (GHZ) measurements. We construct the required resource states using these nonlinearities and incorporate them into the measurement process to realise stabiliser measurements on a surface code. Within this framework, we analyse the loss tolerance of this architecture and demonstrate improved robustness to loss, alongside reduced resource requirements, compared to existing approaches.

\section{GHZ-measurement-based architecture}
In measurement-based quantum computing (MBQC)~\cite{Briegel2009, PhysRevLett.86.5188, Bartolucci2023-tu}, resource states are created and then consumed through a series of measurements that both drive the computation and facilitate error correction. We adopt an MBQC architecture based on small, encoded two-qubit resource states and multipartite entangling measurements~\cite{PhysRevLett.133.050604}. This approach shifts the most demanding aspects of computation away from the resource state generation -- which is comparatively easier for smaller states -- and instead onto the measurement circuits.
We provide a brief introduction to the GHZ-measurement-based architecture in the following section, with more information available in Reference~\cite{PhysRevLett.133.050604}. We use the {cyclic} variant of this architecture which showed better performance than the {minimal} variant. The cyclic architecture incorporates additional encoding redundancy to provide an overcomplete set of measurement outcomes. 

\subsection{Resource states}
A resource state is an entangled state of photons that can be networked to make a logical qubit.
 We use entangled two-qubit states called {two-chains} as our primary resource (see Figure~\ref{fig:measurement_module}(a)). 
 The two-chain state can be described by the stabilisers $Z_1X_2$ and $X_1Z_2$, where $X_i$ ($Z_i$) is the Pauli-X (Pauli-Z) operator acting on the $i$-th qubit.
 This strategic choice is driven by their inherent simplicity, making them significantly easier to generate compared to larger, more complex entangled states.

In this work, we implement the surface code by using four-qubit GHZ measurements which are expressed in terms of Bell state measurements. Each input qubit is encoded in a two-qubit repetition code, with stabilisers $Z_{ia}Z_{ib}$ and the logical operators of the encoded qubits are 
\begin{align}
\begin{split}
Z_i&=Z_{ia}\sim Z_{ib}\\
X_i&=X_{ia}X_{ib},
\end{split}
\end{align}
where $Z_{ia}$ and $Z_{ib}$ are logically equivalent.
The resulting state is called a doubled two-chain (Figure~\ref{fig:measurement_module}(a)) and is described by the stabilisers
\begin{align}
   \mathcal{R} = \{X_{1a}X_{1b}Z_{2a}, Z_{1a}X_{2a}X_{2b},Z_{1a}Z_{1b}, Z_{2a}Z_{2b}\}.
\end{align}

Photon loss is the dominant source of error in optical quantum computing platforms \cite{Bartolucci2023-tu, bartolucci2025comparisonschemeshighlyloss, 10.1093/acrefore/9780190871994.013.84}. 
In photonic architectures, photon loss reduces the success probability of resource-state preparation and leads to erasure events. The former can be improved via multiplexing, while the latter can be mitigated through redundant encoding of the quantum information. We therefore incorporate additional redundancy into our resource states by using the quantum parity code (QPC) resulting in QPC$(n,m)_r$ encoded two-chains, where $r=2$ indicates the doubled chain introduced earlier. Further details on various encoded resource states are provided in Appendix \ref{app:rs-encoding}. 

\subsection{Measurement module}
\begin{figure}
    \includegraphics[width=0.48\textwidth]{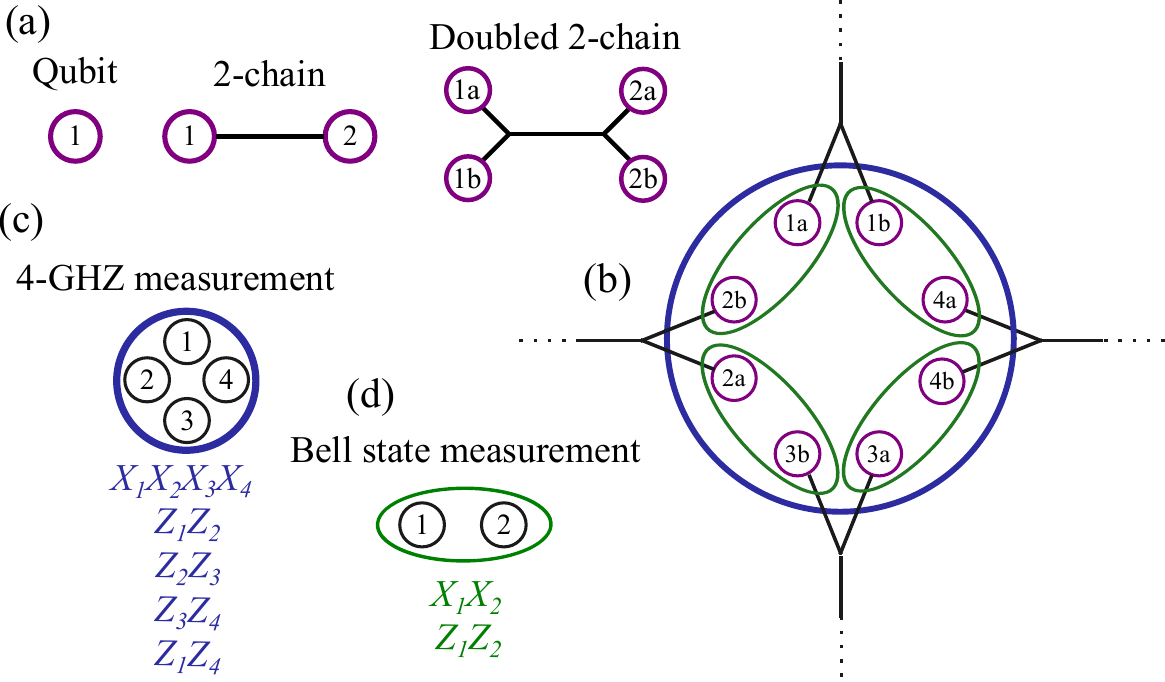}
    \caption{Resource states and measurements in the GHZ-measurement-based architecture. (a) We use encoded doubled two-chains as our resource states and these are sent to a measurement module. (b) The measurement module performs a four-qubit GHZ measurement (blue) on four input qubits. (c) Measurement operators associated with the four-qubit GHZ measurement. Each input qubit is encoded in a two-qubit repetition code (doubled two-chain) to perform the four-qubit GHZ measurement in terms of (d) physical Bell state measurements which measure the operators $X_1X_2$ and $Z_1Z_2$ (green).}
    \label{fig:measurement_module}
\end{figure}

The measurement module (see Figure~\ref{fig:measurement_module}(b)) in our architecture facilitates both quantum computation and fault-tolerant error correction. It measures the Pauli operators
$\{X_1X_2X_3X_4,Z_1Z_2, Z_2Z_3,Z_3Z_4,Z_1Z_4\}$
to identify errors, and reconfigurable measurement bases enable the implementation of Clifford gates on logical qubits.

As mentioned in the preceding section, we decompose the four-qubit GHZ measurements into physical Bell state measurements. 
The measurement outcomes correspond to the eigenvalues of the operators
\begin{align}
\mathcal{M} = \{&X_{1a}X_{2b}, Z_{1a}Z_{2b}, X_{2a}X_{3b}, Z_{2a}Z_{3b},\notag\\
&X_{3a}X_{4b}, Z_{3a}Z_{4b}, X_{4a}X_{1b}, Z_{4a}Z_{1b} \},
\end{align}
which generate the group associated with four-qubit GHZ measurements.
The logical operator ${X_1X_2X_3X_4}$ is given by the product of all $X_{ia}X_{(i+1)b}$ operators, and the logical operator ${Z_iZ_{i+1}} = Z_{ia}Z_{i+1b}$.
The stabiliser group of the resource states and the group of measurement operators determine the check operator group and the corresponding syndrome graph, both of which are discussed in more detail in Appendix \ref{app:syndrome}.

To analyse the loss tolerance of our architecture for a given resource state encoding, we must evaluate the eigenvalue return probability of our measurements. Specifically, for a QPC$(n,m)_r$ encoded two-chain state, we perform QPC$(n,m)$-encoded Bell state measurements (Figure~\ref{fig:measurement_module}(b)).
The eigenvalue return probability quantifies the likelihood that the measurement returns the correct eigenvalue of the target operator and includes the effect of loss, characterised by the loss rate $\eta$. For linear-optical Bell measurements, these are given by 
\begin{align}
\begin{split}
    P(zz^{(n,m)}) & =1-\left[1-\frac{1}{2}\left(1-\gamma\right)^m\right]^n\\
    P(xx^{(n,m)}) & = (1-\gamma^m)^n,
 \end{split}
  \label{eq:rec_prob_lo}
\end{align}
where $\gamma = 1-(1-\eta)^2$, with $\eta =0$ corresponding to the lossless case.
Even in the absence of loss, the probability $P(zz^{(n,m)})$ is less than one, and so in general quantum error correction must accommodate not only for losses but also for these intrinsic measurement-induced failures.  
Note that the eigenvalue return probabilities for the rotated QPC-encoding can be obtained by replacing $zz^{(n,m)}$ with $xx^{(n,m)}$ and vice versa.

\section{Nonlinearities for photonic quantum computing}
Now we turn to the nonlinear architecture, where we introduce an additional component that deterministically implements a nonlinear sign-shift gate~\cite{Knill2001}, by means of a single-photon-level Kerr interaction, i.e., $\ket{1}\!\rightarrow\!\ket{1},\;\ket{2}\!\rightarrow\!-\ket{2}$. Such strong nonlinearities are often accessed by single-emitters strongly coupled to an optical field, for example a single atom in a cavity~\cite{Thomas2022} or a single quantum dot coupled to a waveguide~\cite{nielsen2024programmablenonlinearquantumphotonic}. In these cases, the nonlinear element can also function as a single-photon source~\cite{uppu2020}.

At ORCA, we use atomic ensembles in optical cavities to realise a single-photon nonlinearity. The same physics that underpins the strong Kerr interaction also means the system can be used as a deterministic single-photon source. In principle, this allows for near-deterministic resource state generation and measurements, massively reducing multiplexing requirements and overall resource overhead, while significantly improving the fault-tolerance threshold beyond what is achievable with linear-optical approaches.

\subsection{Nonlinear photonic entangling gate}

\begin{figure}
    \includegraphics[width=0.475\textwidth]{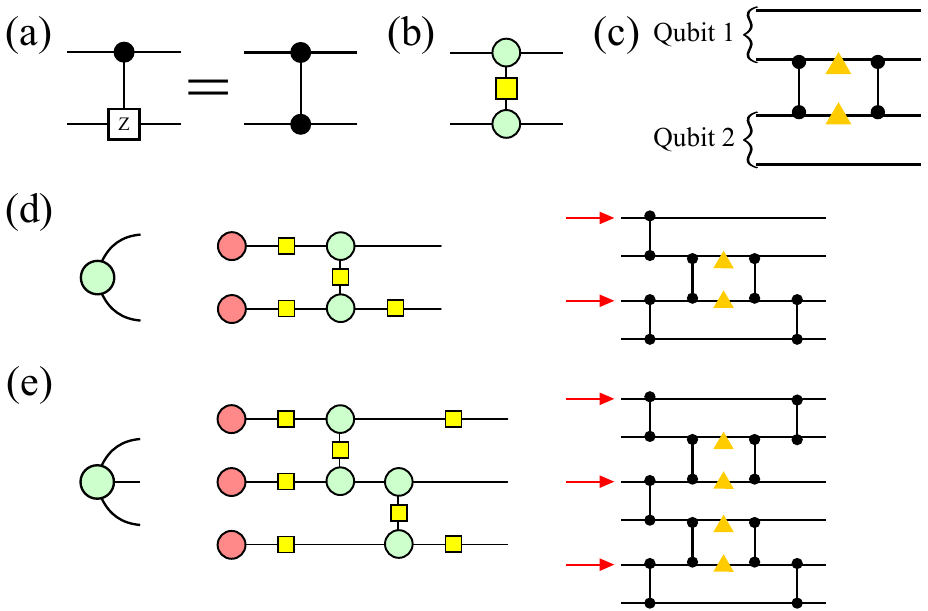}
    \caption{Qubit gates and state preparation using nonlinearities. CZ gate in (a) the qubit picture, (b) using a ZX-diagram, and (c) implemented on dual-rail photonic qubits using balanced beam splitters (vertical black lines) and nonlinearities (yellow triangles). This nonlinear photonic CZ gate can be used to directly generate small seed states, such as (d) Bell states or (e) three-qubit GHZ states, as depicted using equivalent ZX-diagrams and optical circuits.}
    \label{fig:CZ}
\end{figure} 

Given a nonlinearity that induces a relative $\pi$ phase shift between one- and two-photon components, one can realise a nonlinear router that spatially separates these components~ \cite{PickNonlinearRouter,Witthaut_2012}.
That same module also enables direct implementation of a deterministic controlled-Z (CZ) gate in a photonic platform, as illustrated in Figure~\ref{fig:CZ}. This represents a significant advantage over purely linear-optical systems, where such gates are inherently probabilistic and can consume more photons than they output.
The nonlinear module can be used to deterministically generate small entangled seed states \cite{bartolucci2021creationentangledphotonicstates}, such as Bell states and three-qubit GHZ states, with the corresponding optical circuits shown in Figure~\ref{fig:CZ}.

The nonlinearity we consider would enable near-deterministic two-photon gates and so is in principle sufficient for universal gate-based quantum computation. Nevertheless, we adopt a measurement-based approach, motivated by the need to maintain shallow circuit depths to minimise photon loss, which is crucial for achieving fault tolerance for photonic systems.
In our architecture, the single-photon nonlinearity is used both in the resource-state generator, where it enables the generation of entangled states, and for near-deterministic operations in the measurement module.
\subsection{Resource state generator} 
\begin{figure}
    \includegraphics[width=0.48\textwidth]{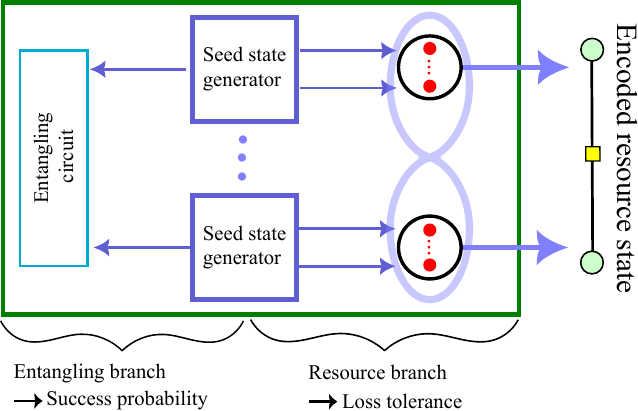}
    \caption{
    Schematic of a resource state generator. A subset of photons from each seed state is routed to an entangling circuit. Upon successful detection of photons within this circuit, the photons in the resource branch are projected into an encoded 2-chain state. Losses in the entangling branch decrease the success probability of resource state generation and losses in the resource branch directly impact the loss threshold relevant to computation and error correction.}
    \label{fig:rsg}
\end{figure} 

\begin{figure}
    \includegraphics[width=0.45\textwidth]{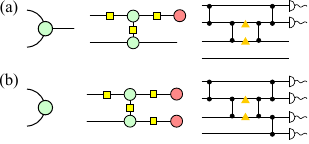}
    \caption{Entangling gates based on the nonlinear photonic CZ gate. (a) two-to-one-qubit entangling gate and (b) two-qubit entangling gate (Bell state measurement). These are used in the entangling circuit for resource state generation. The circuit shown in (b) is also used in the measurement module.}
    \label{fig:fusion}
\end{figure}

The resource state generation process begins with the deterministic generation of single photons, which are routed into nonlinear seed-state generators to produce small entangled states (Figure~\ref{fig:CZ}(d,e)). One photon from each seed-state generator is routed to an entangling circuit which projects the surviving seed-state photons into the desired resource state (Figure~\ref{fig:rsg}). Within the entangling circuit, the photons undergo both partial and full $n$-qubit entangling operations. Example circuits for the case of $n=2$ are shown in Figure~\ref{fig:fusion}.

Photons propagating through entangling circuits may experience loss, but such loss is heralded -- if an insufficient number of photons are detected within the entangling circuit, the generation attempt is discarded. As a result, loss is the only mechanism that reduces the success probability of the nonlinear entangling circuit. Consequently, the multiplexing requirements are significantly reduced compared to linear-optical approaches, where multiplexing must compensate for both loss and the probabilistic nature of state generation and entangling measurements. We determine that just a single stage of multiplexing is sufficient to make the overall nonlinear resource state generation near-deterministic, compared to some estimates of six stages for linear-optical architectures~\cite{wein2024minimizingresourceoverheadfusionbased}.

The output photons forming the resource state will also be subject to loss. However, these photons interact only with the seed-state generator before being measured for computation and error correction and so the associated loss is correspondingly low.

\subsection{Improved performance with nonlinearities}
Optical nonlinearities are incorporated into the implementation of the four-qubit GHZ measurements, enabling near-deterministic performance. For Shor-encoded QPC states, the eigenvalue return probabilities are
\begin{align}
    \begin{split}
    P(zz^{(n,m)}) & =1-\left[1-(1-\gamma)^m\right]^n\\
    P(xx^{(n,m)}) & = (1-\gamma^m)^n,
    \end{split}
\end{align}
where the first probability now exceeds the value for the linear-optical approach given in Eq.~(\ref{eq:rec_prob_lo}).
The error correction scheme no longer has to account for failures of the entangling measurement arising from the probabilistic nature of linear-optical circuits. This significantly enhances the loss tolerance of our architecture, as we discuss in Section~\ref{sec:ft}.

\subsection{Networking}
Networking logical qubits is essential for running large-scale algorithms and typically presents a significant challenge for hardware platforms. In matter-based qubit platforms, logical qubits located on different chips cannot be easily interconnected \cite{Storz2023}. Establishing such connections requires the development of optical interfaces that enable correlations between qubits through entangling operations, such as Bell-state measurements \cite{riedel2025scalablephotonicquantuminterconnect, deGliniasty2024spinopticalquantum,inc2024distributedquantumcomputingsilicon}. Consequently, these platforms must introduce entirely new technologies to support logical multi-qubit gates.
In contrast, photonic quantum computing offers a more natural way to network qubits because the same techniques used to generate correlations between physical qubits can be extended to logical qubits. However, in both linear-optic and matter-based platforms, networking will be constrained by the probabilistic nature of linear-optical entangling measurements.
Our nonlinear platform offers a substantial advantage over approaches that rely on linear optics for networking qubits. We can create entanglement deterministically using the CZ gate introduced in Figure~\ref{fig:CZ} and the entangling measurements in Figure~\ref{fig:fusion}(b). Networking in our platform uses the same technology that is being developed for the creation of our resource states and measurement modules.

\section{Fault tolerance}
\label{sec:ft}
We develop a universal, fault-tolerant photonic architecture for quantum computing. We use the foliated rotated surface code \cite{PhysRevLett.117.070501} as our quantum error correction scheme to mitigate the impact of errors.
It is an adaptation of the standard surface code tailored for MBQC, where the two-dimensional code is transformed into a three-dimensional structure, enabling its use in photonic quantum architectures.
This code is constructed by stacking multiple layers of the surface code along the temporal axis, resulting in a large-scale fault-tolerant structure known as the Raussendorf-Harrington-Goyal (RHG) lattice \cite{Raussendorf_2007}.

Loss is the main error mechanism for photonic quantum computers and is flagged by failure to detect the expected number of photons. As outlined earlier, we use additional redundancy in our resource states and shallow optical circuits -- where each photon interacts with only a few components before being detected -- to mitigate loss.\\

\begin{figure}
    \includegraphics[width=0.48\textwidth]{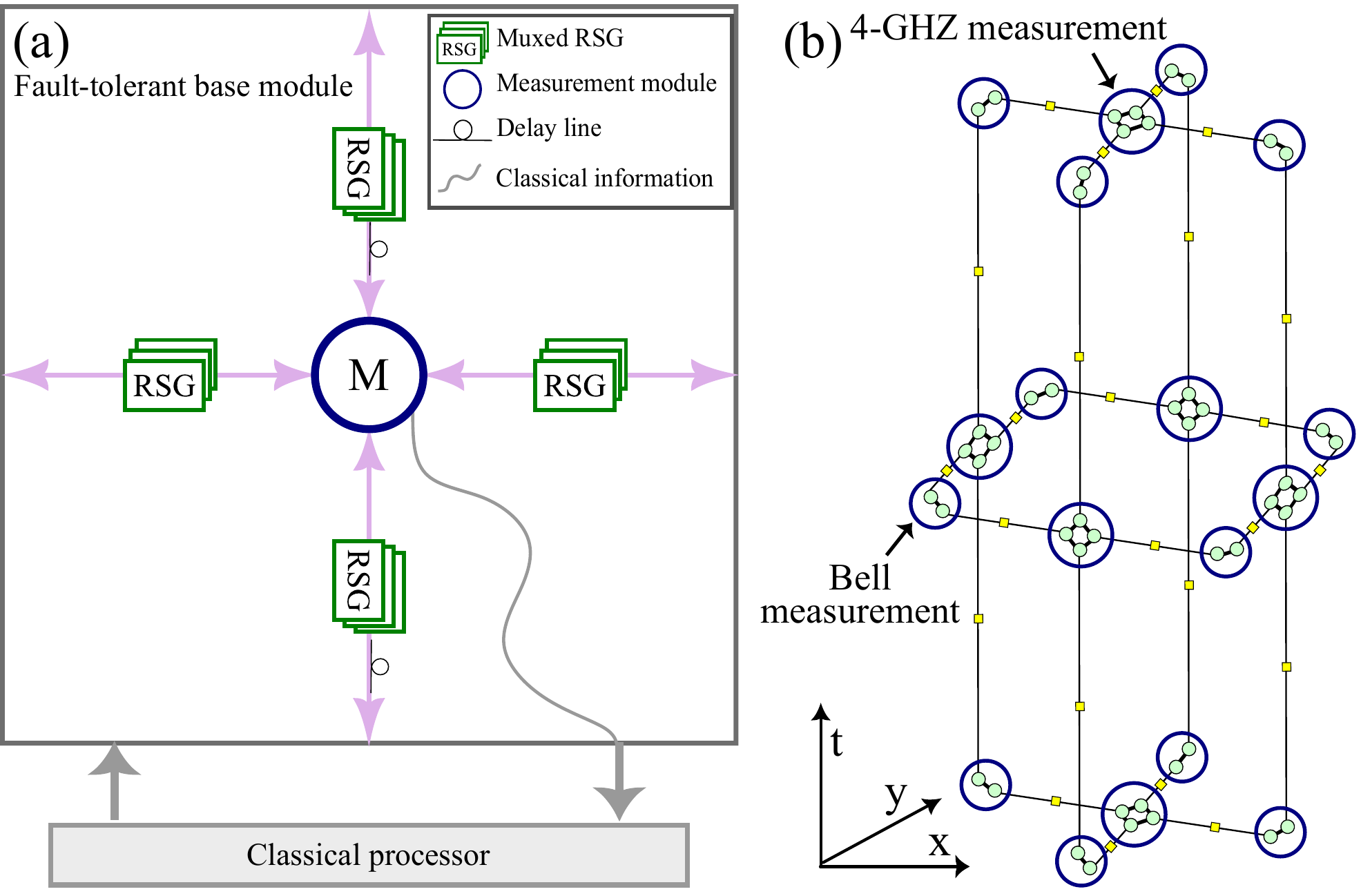}
    \caption{(a) Fault-tolerant base module comprising resource state generators (RSGs) and a measurement module. The RSGs are multiplexed to make the generation near-deterministic. Classical information from the measurement module is routed to a classical processor which handles decoding, Pauli frame tracking, feed-forward and change of measurement basis to implement logic. (b) Copies of this base module can be connected to generate the RHG lattice, with a representative cell shown here. Encoded two-chains are combined using entangling operations (circles around nodes connected by thicker lines) which are four-qubit GHZ measurements in the bulk of the lattice and measurements on smaller numbers of qubits (e.g., Bell measurements) at the boundaries.}
    \label{fig:RHG}
\end{figure}

We construct the RHG lattice by using copies of a base module that creates multiplexed resource states and routes them to the appropriate measurements (Figure~\ref{fig:RHG}(a)). A representative cell of the RHG lattice, constructed from two-chain states and multiqubit measurements, is shown in Figure~\ref{fig:RHG}(b). A classical processor takes classical information from the measurement module and performs decoding, configures switch settings to implement logic, and keeps track of the Pauli frame, which records accumulated Pauli errors that do not require physical correction during computation.\\

The loss threshold describes the maximum loss per photon that can be tolerated while protecting the logical qubit.  We determine loss thresholds through Monte Carlo simulations that check when qubit losses span opposing boundaries of the surface code on either the primal or dual lattice, with each lattice treated independently, resulting in a logical loss. For each resource state,  we run 50k Monte Carlo samples and check for a logical loss to estimate the logical error rate. The simulated surface code has dimension $d\times d\times (2d+1)$, and we scan the loss rate $\eta$ against the logical error rate for code distances $d\in \{11,13,15\}$. The results are shown for different encodings of our resource state in Table \ref{tab:loss}. 

\begin{table}[t]
    \centering
    \begin{tabular}{l|c |c }

       \thead{Encoded two-chain}&\thead{Number of\\ photons}& \thead{Loss per photon \\threshold (\%)}\\
        \hline
        \hline
      Doubled ($r=2$) &4&3.1 \\ 
      Shor encoded QPC$(1,2)_2$ &8&5.3\\ Shor encoded QPC$(2,2)_2$&16&8.7    \\ Rotated QPC$(4,2)_2$&32&11.5 \\ Shor encoded QPC$(6,4)_2$&96&15.1 
    \end{tabular}
    \caption{Single-photon loss thresholds as a function of the number of photons in the resource state for our nonlinear architecture. The resource states are QPC$(n,m)_r$ encoded two-chain states.}
    \label{tab:loss}
\end{table}
\begin{figure}
    \includegraphics[width=0.48\textwidth]{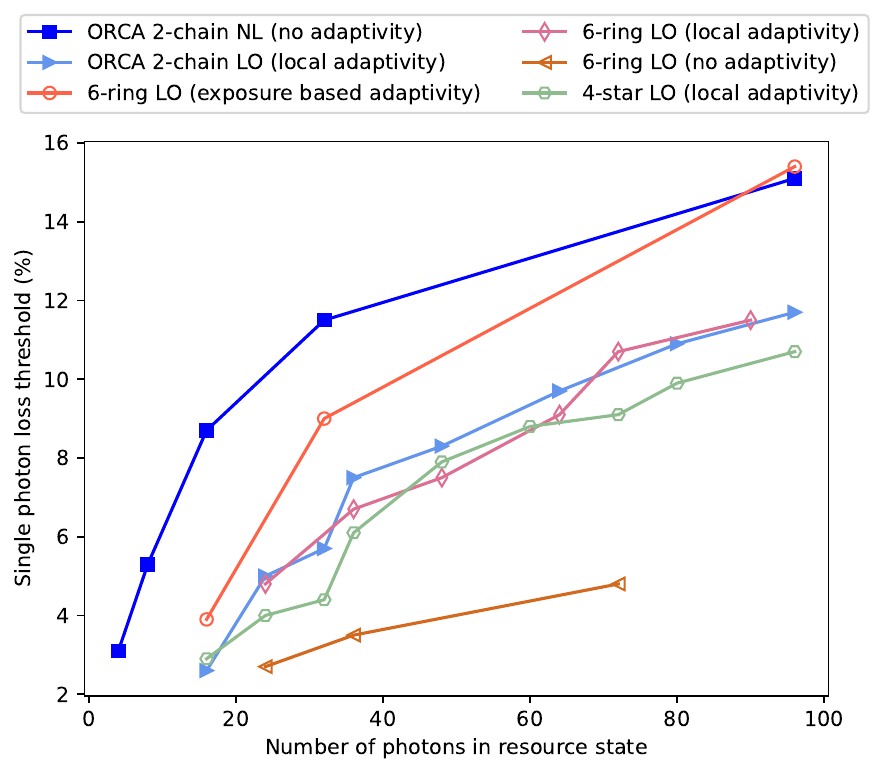}
    \caption{Comparison of single-photon loss thresholds for nonlinear (NL) and linear-optical (LO) architectures. Adaptivity is not used for our NL thresholds (closed squares) but is used for LO thresholds to enhance loss tolerance. The closed triangles denote the GHZ-measurement-based architecture~\cite{PhysRevLett.133.050604} without nonlinearities, while all other LO values correspond to fusion-based architectures employing six-ring (red) and four-star (green) resource states~\cite{Bartolucci2023-tu,bartolucci2025comparisonschemeshighlyloss}.}
    \label{fig:loss_threshold}
\end{figure} 
The loss thresholds of our nonlinear architecture are calculated without making use of advanced techniques, such as adaptivity or active error mitigation, yet still remain competitive with linear-optical approaches which do incorporate such techniques, as shown in Figure~\ref{fig:loss_threshold}. We anticipate that the threshold for our nonlinear approach can be further improved by using adaptivity.
Our architecture provides the required ingredients for universal fault-tolerant quantum computation within the measurement-based framework. In particular, universality is achieved in the standard way via magic state injection for non-Clifford operations \cite{PhysRevA.71.022316}. The architecture can also implement alternative error-correcting codes by adapting the number of qubits involved in the multipartite measurements and modifying their routing across the networked modules.

\section{Implementation}
Our architecture can be optimised to trade off hardware efficiencies against resource overheads for generating and entangling photons. We choose the 32-photon resource state with 11.5\% loss threshold and find that sub-threshold operation is compatible with projected efficiencies for our proprietary components, such as integrated
photonic optical switches and optical nonlinearities.
Photonics has a significant scaling advantage over other platforms: The number of photonic qubits can be increased  by leveraging the time domain through optical-fibre delays \cite{bombin2021interleavingmodulararchitecturesfaulttolerant}. Each resource state generator can be reused to produce many resource states in the same optical fibre. This is in contrast to matter-based systems, where scaling necessarily means more physical qubits and a larger footprint. For a single logical qubit, we estimate our architecture requires fewer than 10k nonlinear single-photon sources, which is several orders of magnitude less than numbers reported for linear-optical architectures~\cite{AghaeeRad2025,wein2024minimizingresourceoverheadfusionbased}.

The biggest footprint factor for purely linear-optical designs comes from the cryogenic cooling of many modules, driven by the need to tightly co-integrate components with superconducting photon detectors to implement low-latency multiplexing logic \cite{AghaeeRad2025,alexander2024manufacturableplatformphotonicquantum}. These systems can comprise substantial custom cryoplants with building-size footprints. In contrast, the minimal multiplexing requirements of our nonlinear architecture allow us to decouple the cryogenic requirements of photon detection from all other optical circuitry: Our photonic integrated circuits, switch networks, electronics, nonlinearities and delay lines do not need to be engineered for cryogenic operation. The system therefore operates mainly at room temperature and only a modest number of detectors need to be cooled to realise a single logical qubit, putting our needs within reach of commercial solutions.

\section{Conclusion}
In this work, we have introduced a universal fault-tolerant photonic measurement-based quantum computing architecture based on single-photon nonlinearities. The architecture enables the near-deterministic construction of entangled resource states and measurement operations that implement error correction and underpin the computational framework. This results in significantly improved resource efficiency and enhanced loss tolerance compared to existing approaches.

Our analysis focuses on photon loss as the dominant error mechanism. Other error sources, such as photon distinguishability~\cite{steinmetz2024simulating} and multiphoton contamination, can be incorporated into our analysis and, to leading order, are expected to introduce additional erasure errors and Pauli-type errors. We have assumed ideal $\pi$-nonlinearities in this study, and future work will incorporate more sophisticated models to accurately describe the performance of the practical implementations we are developing. Our architecture is also compatible with non-$\pi$ nonlinearities which provides additional flexibility in the implementation.

Scaling remains one of the central challenges facing the construction of universal, fault-tolerant quantum computers. Our nonlinear photonic architecture significantly reduces multiplexing overheads and physical footprint, enhances loss tolerance, and lowers component performance requirements. It therefore enables the development of scalable systems that leverage the inherent modularity of photonic interconnects, while avoiding both the extensive high-vacuum and cryogenic infrastructure needed to sustain spin coherence, and the impractical spatial footprints of warehouse-scale linear-optical systems. Nonlinear photonic architectures provide a route to dramatically reducing the size and timeline for fault-tolerant quantum computing.\\

\section{Acknowledgements}
The authors thank Alex Neville for helpful input on many aspects of the nonlinear architecture, including nonlinear resource state generation and encodings, calculations of loss tolerances, and the use of ZX-calculus to simplify the architecture. We also thank Peter Mosley, Omkar Srikrishna, Richard Tatham and our colleagues at ORCA Computing for useful discussions and comments.\\

\section{Appendices}

\appendix

\section{Resource state encoding}
\label{app:rs-encoding}
The resource states used in our quantum computing architecture are minimal in size, consisting of two-qubit graph states, referred to as two-chain states. To implement GHZ measurements using Bell measurements, we apply a two-qubit repetition code to each qubit, an approach we refer to as doubling ($r=2$).
To further enhance the loss tolerance of the system, we encode each node in the doubled two-chain resource state in a QPC$(n,m)$ code.
A QPC$(n,m)$ encoding refers to a quantum parity check code, which can be viewed as a generalisation of the Shor code \cite{bacon2006quantumerrorcorrectingsubsystem}. Here, 
$m$ denotes the number of physical qubits used to encode a single block-level qubit, and 
$n$ specifies the number of such blocks in the code
\begin{align}
    \ket{0^{(n,m)}}_{Shor} &= \left(\frac{1}{\sqrt{2}}\left(\ket{0}^{\otimes m}+\ket{1}^{\otimes m}\right)\right)^{\otimes n}\\
       \ket{1^{(n,m)}}_{Shor} &= \left(\frac{1}{\sqrt{2}}\left(\ket{0}^{\otimes m}-\ket{1}^{\otimes m}\right)\right)^{\otimes n}.
\end{align}

\begin{figure}
    \includegraphics[width=0.48\textwidth]{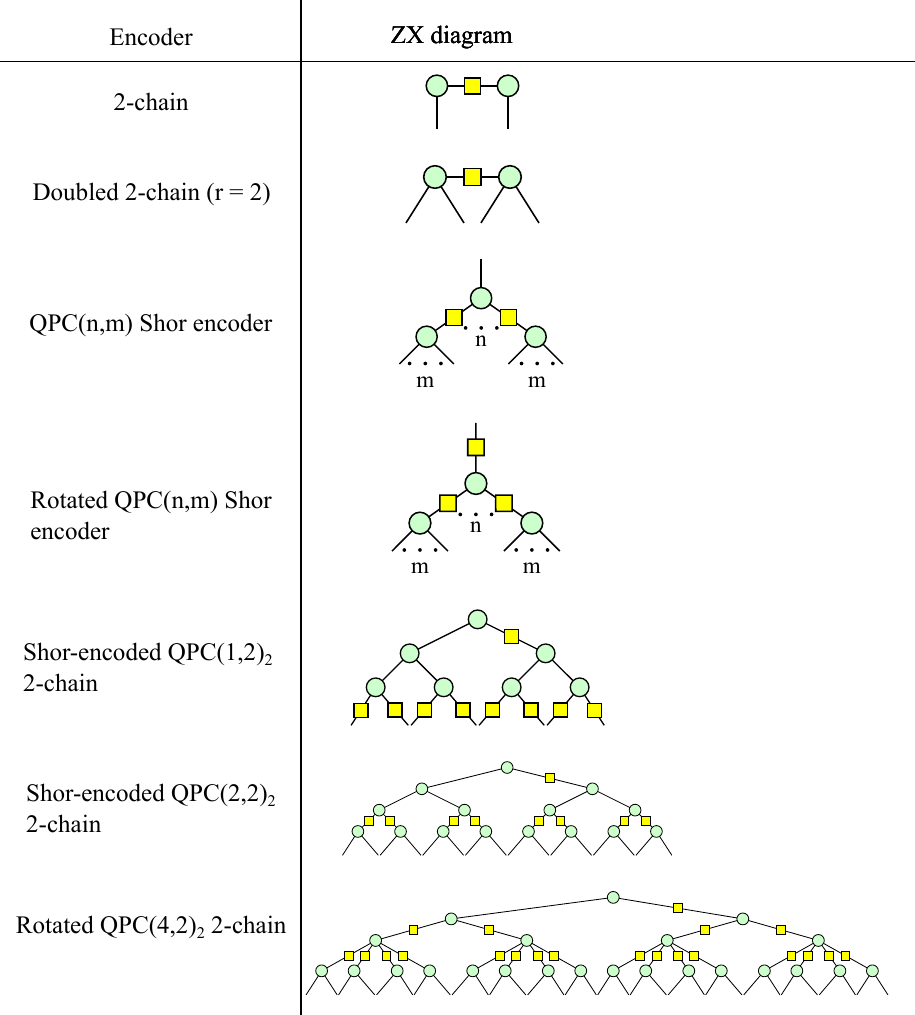}
    \caption{The resource state in our architecture is a two-chain state. The two-chain state is first encoded in a repetition code which means we end up with a doubled two-chain ($r=2$). Each qubit in the double two-chain can be further encoded in a QPC$(n,m)$ code to improve the loss tolerance of the state.}   \label{fig:zx_encoding}
\end{figure}

We adopt the following convention for the rotated QPC encoding
\begin{align}
    \ket{+^{(n,m)}}_{rot} &=\ket{0^{(n,m)}}_{Shor} \\
     \ket{-^{(n,m)}}_{rot} &=\ket{1^{(n,m)}}_{Shor}. 
\end{align}
As a result, our resource state becomes a QPC$(n,m)_r$-encoded two-chain. We evaluate the threshold performance for several such encoded two-chain states. The corresponding encoding circuits (or encoders), up to the QPC$(4,2)_2$-encoded two-chain, are provided in Figure~\ref{fig:zx_encoding}, using the framework developed in \cite{PhysRevA.110.032402}.

\section{Syndrome graph}
\begin{figure}
   \includegraphics[width=0.48\textwidth]{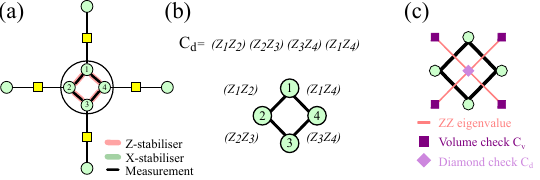}
   \caption{Diamond check operator C$_{\rm{d}}$ for a 4-qubit GHZ measurement. (a) ZX-diagram representation of four two-chain resource states, where one qubit from each resource state participates in the four-qubit GHZ measurement, implemented via Bell measurements. Pauli webs \cite{Bombin2024unifyingflavorsof} illustrate the emergence of the diamond check operator, where thicker lines indicate a measurement outcome. (b) Explicit depiction of the diamond check operator associated with the four-qubit GHZ measurement. (c) Syndrome graph corresponding to the four-qubit GHZ measurement constructed from the $ZZ$ measurement outcomes.}
   \label{fig:4ghz_syndrome}
\end{figure} 
\begin{figure}
   \includegraphics[width=0.44\textwidth]{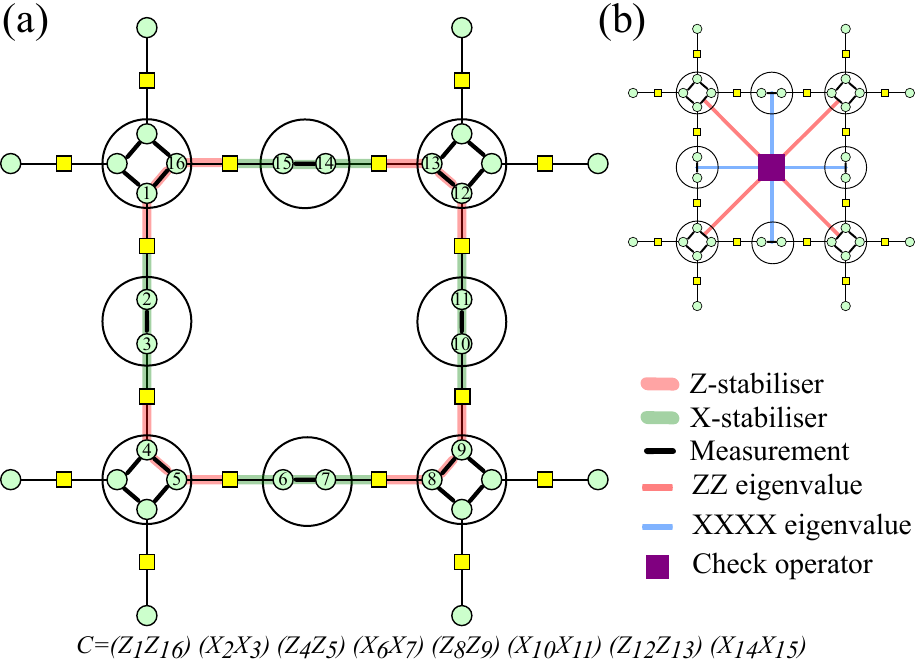}
   \caption{(a) Example of a 2D slice of the RHG lattice (foliated surface code) represented in
   ZX-calculus. Check operators are illustrated using Pauli webs. (b) Corresponding syndrome graph.}
   \label{fig:2Dsquare_syndrome}
\end{figure} 
\begin{figure}
   \includegraphics[width=0.44\textwidth]{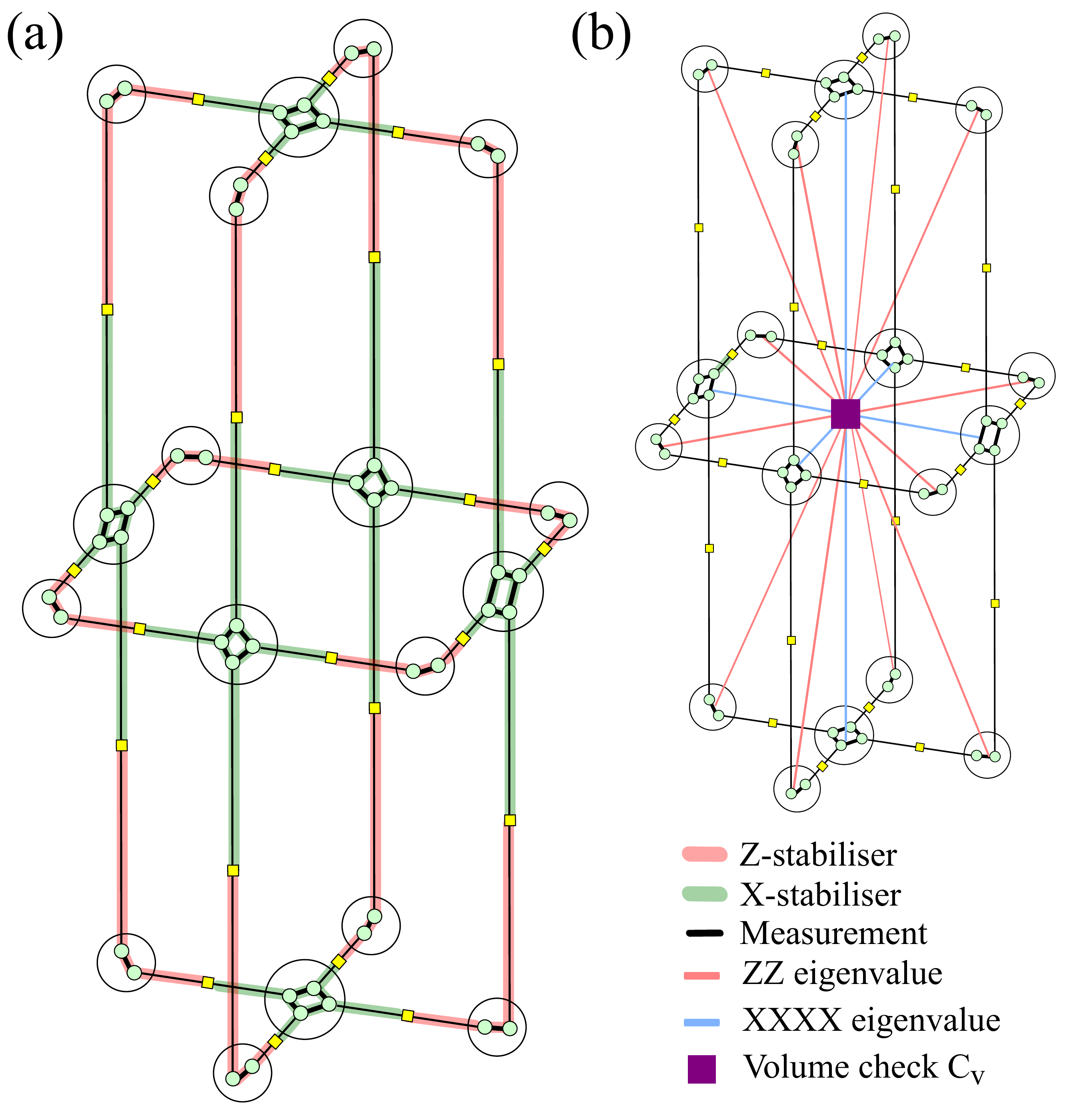}
   \caption{(a) Unit cell of the RHG lattice represented using ZX-diagrams and with Pauli web overlay. (b) Corresponding syndrome graph and volume check operator.}
   \label{fig:unit_cell_syndrome}
\end{figure} 
\label{app:syndrome}

In our measurement-based quantum computing framework, the error correction code is constructed using resource states that define the state of the system through a stabiliser group $\mathcal R$ \cite{gottesman1997stabilizer}.

This group $\mathcal{R}$ comprises the stabilisers associated with all resource states involved in the computation. During the execution of quantum operations and error correction through measurements, a measurement group $\mathcal M$ is generated. The set of check operators $\mathcal{C}$ is then defined as the group $\mathcal C = \mathcal R \cap \mathcal M$, which serves to detect specific errors occurring during the computation \cite{Raussendorf_2007,bartolucci2025comparisonschemeshighlyloss, PhysRevResearch2_033305, PhysRevLett.133.050604}. Each error is associated with precisely two violated parity check operators \cite{Raussendorf_2007, Bartolucci2023-tu}, and the collection of these violations constitutes the error syndrome.

Detectable errors are those that anticommute with at least one stabiliser in the measurement group, resulting in a change in the eigenvalue (parity) of two check operators. In contrast, undetectable errors commute with all stabilisers in the measurement group and therefore leave the eigenvalues of all check operators unchanged. As a result, these errors do not produce a detectable syndrome and cannot be identified through measurement alone.

To visualise these syndromes we use a graphical representation of the observed error patterns called the syndrome graph. In this graph, vertices correspond to check operators, while edges represent individual measurement outcomes. The syndrome graph provides a powerful framework for identifying and correcting errors in quantum error correction codes, enabling a systematic analysis of error patterns and efficient recovery strategies.

If an even number of errors occur at a given check operator, their combined effect commutes with the check operator, and thus the overall parity remains unchanged. Consequently, no violation is detected at that check operator. This phenomenon allows error chains to form when multiple correlated errors collectively preserve the parity at intermediate check operators, thus only triggering violations at the endpoints. These chains manifest in the syndrome graph as paths connecting pairs of check operators with flipped parity which represent the endpoints of the error syndrome.

Stabiliser generators and check operators of a quantum state can be identified using ZX-calculus  \cite{Coecke_2011, PhysRevA.110.032402}, together with Pauli webs \cite{Bombin2024unifyingflavorsof}, which are graphical decorations on the ZX-diagram that reveal the underlying stabiliser structure of the state. Check operators correspond to Pauli webs that have support exclusively on internal nodes, with no connections to external legs of the ZX-diagram. We use Pauli webs to present three examples of check operators and their associated syndrome graphs: a single 4-qubit GHZ measurement (Figure~\ref{fig:4ghz_syndrome}), a two-dimensional slice of the surface code (Figure~\ref{fig:2Dsquare_syndrome}), and a full unit cell of the RHG lattice (Figure~\ref{fig:unit_cell_syndrome}). The primal syndrome graph depicts red-colored edges corresponding to $ZZ$-type outcomes, and blue-colored edges correspond to $\prod_i X_i$-type outcomes.
For the dual lattice, the complementary set of measurement outcomes -- those not used in the primal lattice -- are used to extract the syndrome. 
Photon loss in the resource branch or the measurement module results in the erasure of outcomes from the 4-GHZ measurements. When such eigenvalues are missing, the standard construction of check operators is no longer feasible.
To enable continued error detection in the presence of these erasures, we construct supercheck operators \cite{PhysRevA.97.030301,PhysRevLett.105.200502} by combining the two affected check operators. The resulting supercheck is independent of the missing measurement outcome. Logical loss occurs when the erasures span the lattice in such a way that they connect opposite boundaries. While we currently focus only on loss errors, the deformed lattice constructed from checks and superchecks is used to decode Pauli errors using standard decoders such as minimum-weight perfect matching \cite{higgott2021pymatchingpythonpackagedecoding} or union-find decoding \cite{Delfosse2021almostlineartime}.


\begin{thebibliography}{59}%
\makeatletter
\providecommand \@ifxundefined [1]{%
 \@ifx{#1\undefined}
}%
\providecommand \@ifnum [1]{%
 \ifnum #1\expandafter \@firstoftwo
 \else \expandafter \@secondoftwo
 \fi
}%
\providecommand \@ifx [1]{%
 \ifx #1\expandafter \@firstoftwo
 \else \expandafter \@secondoftwo
 \fi
}%
\providecommand \natexlab [1]{#1}%
\providecommand \enquote  [1]{``#1''}%
\providecommand \bibnamefont  [1]{#1}%
\providecommand \bibfnamefont [1]{#1}%
\providecommand \citenamefont [1]{#1}%
\providecommand \href@noop [0]{\@secondoftwo}%
\providecommand \href [0]{\begingroup \@sanitize@url \@href}%
\providecommand \@href[1]{\@@startlink{#1}\@@href}%
\providecommand \@@href[1]{\endgroup#1\@@endlink}%
\providecommand \@sanitize@url [0]{\catcode `\\12\catcode `\$12\catcode `\&12\catcode `\#12\catcode `\^12\catcode `\_12\catcode `\%12\relax}%
\providecommand \@@startlink[1]{}%
\providecommand \@@endlink[0]{}%
\providecommand \url  [0]{\begingroup\@sanitize@url \@url }%
\providecommand \@url [1]{\endgroup\@href {#1}{\urlprefix }}%
\providecommand \urlprefix  [0]{URL }%
\providecommand \Eprint [0]{\href }%
\providecommand \doibase [0]{https://doi.org/}%
\providecommand \selectlanguage [0]{\@gobble}%
\providecommand \bibinfo  [0]{\@secondoftwo}%
\providecommand \bibfield  [0]{\@secondoftwo}%
\providecommand \translation [1]{[#1]}%
\providecommand \BibitemOpen [0]{}%
\providecommand \bibitemStop [0]{}%
\providecommand \bibitemNoStop [0]{.\EOS\space}%
\providecommand \EOS [0]{\spacefactor3000\relax}%
\providecommand \BibitemShut  [1]{\csname bibitem#1\endcsname}%
\let\auto@bib@innerbib\@empty
\bibitem [{\citenamefont {Breuckmann}\ and\ \citenamefont {Eberhardt}(2021)}]{PRXQuantum.2.040101}%
  \BibitemOpen
  \bibfield  {author} {\bibinfo {author} {\bibfnamefont {N.~P.}\ \bibnamefont {Breuckmann}}\ and\ \bibinfo {author} {\bibfnamefont {J.~N.}\ \bibnamefont {Eberhardt}},\ }\href {https://doi.org/10.1103/PRXQuantum.2.040101} {\bibfield  {journal} {\bibinfo  {journal} {PRX Quantum}\ }\textbf {\bibinfo {volume} {2}},\ \bibinfo {pages} {040101} (\bibinfo {year} {2021})}\BibitemShut {NoStop}%
\bibitem [{\citenamefont {Komoto}\ and\ \citenamefont {Kasai}(2025)}]{komoto2024quantum}%
  \BibitemOpen
  \bibfield  {author} {\bibinfo {author} {\bibfnamefont {D.}~\bibnamefont {Komoto}}\ and\ \bibinfo {author} {\bibfnamefont {K.}~\bibnamefont {Kasai}},\ }\bibfield  {journal} {\bibinfo  {journal} {npj Quantum Information}\ }\textbf {\bibinfo {volume} {11}},\ \href {https://doi.org/10.1038/s41534-025-01090-1} {10.1038/s41534-025-01090-1} (\bibinfo {year} {2025})\BibitemShut {NoStop}%
\bibitem [{\citenamefont {{Google Quantum AI}}\ and\ \citenamefont {{Collaborators}}(2025)}]{Acharya2025}%
  \BibitemOpen
  \bibfield  {author} {\bibinfo {author} {\bibnamefont {{Google Quantum AI}}}\ and\ \bibinfo {author} {\bibnamefont {{Collaborators}}},\ }\href {https://doi.org/10.1038/s41586-024-08449-y} {\bibfield  {journal} {\bibinfo  {journal} {Nature}\ }\textbf {\bibinfo {volume} {638}},\ \bibinfo {pages} {920} (\bibinfo {year} {2025})}\BibitemShut {NoStop}%
\bibitem [{\citenamefont {Zhao}\ \emph {et~al.}(2025)\citenamefont {Zhao}, \citenamefont {Singh}, \citenamefont {Singh}, \citenamefont {Thoreen}, \citenamefont {DeAngelo}, \citenamefont {Dominguez}, \citenamefont {Leenheer}, \citenamefont {Peyskens}, \citenamefont {Lukin}, \citenamefont {Englund}, \citenamefont {Eichenfield}, \citenamefont {Gemelke},\ and\ \citenamefont {Wan}}]{zhao2025integratedphotonicsplatformhighspeed}%
  \BibitemOpen
  \bibfield  {author} {\bibinfo {author} {\bibfnamefont {M.}~\bibnamefont {Zhao}}, \bibinfo {author} {\bibfnamefont {M.}~\bibnamefont {Singh}}, \bibinfo {author} {\bibfnamefont {A.}~\bibnamefont {Singh}}, \bibinfo {author} {\bibfnamefont {H.}~\bibnamefont {Thoreen}}, \bibinfo {author} {\bibfnamefont {R.~J.}\ \bibnamefont {DeAngelo}}, \bibinfo {author} {\bibfnamefont {D.}~\bibnamefont {Dominguez}}, \bibinfo {author} {\bibfnamefont {A.}~\bibnamefont {Leenheer}}, \bibinfo {author} {\bibfnamefont {F.}~\bibnamefont {Peyskens}}, \bibinfo {author} {\bibfnamefont {A.}~\bibnamefont {Lukin}}, \bibinfo {author} {\bibfnamefont {D.}~\bibnamefont {Englund}}, \bibinfo {author} {\bibfnamefont {M.}~\bibnamefont {Eichenfield}}, \bibinfo {author} {\bibfnamefont {N.}~\bibnamefont {Gemelke}},\ and\ \bibinfo {author} {\bibfnamefont {N.~H.}\ \bibnamefont {Wan}},\ }\href {https://arxiv.org/abs/2508.09920} {\bibfield  {journal} {\bibinfo  {journal} {arXiv:2508.09920}\ } (\bibinfo {year} {2025})}\BibitemShut {NoStop}%
\bibitem [{\citenamefont {Delaney}\ \emph {et~al.}(2024)\citenamefont {Delaney}, \citenamefont {Sletten}, \citenamefont {Cich}, \citenamefont {Estey}, \citenamefont {Fabrikant}, \citenamefont {Hayes}, \citenamefont {Hoffman}, \citenamefont {Hostetter}, \citenamefont {Langer}, \citenamefont {Moses}, \citenamefont {Perry}, \citenamefont {Peterson}, \citenamefont {Schaffer}, \citenamefont {Volin}, \citenamefont {Vittorini},\ and\ \citenamefont {Burton}}]{PhysRevX.14.041028}%
  \BibitemOpen
  \bibfield  {author} {\bibinfo {author} {\bibfnamefont {R.~D.}\ \bibnamefont {Delaney}}, \bibinfo {author} {\bibfnamefont {L.~R.}\ \bibnamefont {Sletten}}, \bibinfo {author} {\bibfnamefont {M.~J.}\ \bibnamefont {Cich}}, \bibinfo {author} {\bibfnamefont {B.}~\bibnamefont {Estey}}, \bibinfo {author} {\bibfnamefont {M.~I.}\ \bibnamefont {Fabrikant}}, \bibinfo {author} {\bibfnamefont {D.}~\bibnamefont {Hayes}}, \bibinfo {author} {\bibfnamefont {I.~M.}\ \bibnamefont {Hoffman}}, \bibinfo {author} {\bibfnamefont {J.}~\bibnamefont {Hostetter}}, \bibinfo {author} {\bibfnamefont {C.}~\bibnamefont {Langer}}, \bibinfo {author} {\bibfnamefont {S.~A.}\ \bibnamefont {Moses}}, \bibinfo {author} {\bibfnamefont {A.~R.}\ \bibnamefont {Perry}}, \bibinfo {author} {\bibfnamefont {T.~A.}\ \bibnamefont {Peterson}}, \bibinfo {author} {\bibfnamefont {A.}~\bibnamefont {Schaffer}}, \bibinfo {author} {\bibfnamefont {C.}~\bibnamefont {Volin}}, \bibinfo {author} {\bibfnamefont {G.}~\bibnamefont {Vittorini}},\ and\ \bibinfo {author}
  {\bibfnamefont {W.~C.}\ \bibnamefont {Burton}},\ }\href {https://doi.org/10.1103/PhysRevX.14.041028} {\bibfield  {journal} {\bibinfo  {journal} {Physical Review X}\ }\textbf {\bibinfo {volume} {14}},\ \bibinfo {pages} {041028} (\bibinfo {year} {2024})}\BibitemShut {NoStop}%
\bibitem [{\citenamefont {Carrera~Vazquez}\ \emph {et~al.}(2024)\citenamefont {Carrera~Vazquez}, \citenamefont {Tornow}, \citenamefont {Rist{\`e}}, \citenamefont {Woerner}, \citenamefont {Takita},\ and\ \citenamefont {Egger}}]{CarreraVazquez2024}%
  \BibitemOpen
  \bibfield  {author} {\bibinfo {author} {\bibfnamefont {A.}~\bibnamefont {Carrera~Vazquez}}, \bibinfo {author} {\bibfnamefont {C.}~\bibnamefont {Tornow}}, \bibinfo {author} {\bibfnamefont {D.}~\bibnamefont {Rist{\`e}}}, \bibinfo {author} {\bibfnamefont {S.}~\bibnamefont {Woerner}}, \bibinfo {author} {\bibfnamefont {M.}~\bibnamefont {Takita}},\ and\ \bibinfo {author} {\bibfnamefont {D.~J.}\ \bibnamefont {Egger}},\ }\href {https://doi.org/10.1038/s41586-024-08178-2} {\bibfield  {journal} {\bibinfo  {journal} {Nature}\ }\textbf {\bibinfo {volume} {636}},\ \bibinfo {pages} {75} (\bibinfo {year} {2024})}\BibitemShut {NoStop}%
\bibitem [{\citenamefont {Gidney}(2025)}]{gidney2025factor2048bitrsa}%
  \BibitemOpen
  \bibfield  {author} {\bibinfo {author} {\bibfnamefont {C.}~\bibnamefont {Gidney}},\ }\href {https://arxiv.org/abs/2505.15917} {\bibfield  {journal} {\bibinfo  {journal} {arXiv:2505.15917}\ } (\bibinfo {year} {2025})}\BibitemShut {NoStop}%
\bibitem [{\citenamefont {Mohseni}\ \emph {et~al.}(2025)\citenamefont {Mohseni}, \citenamefont {Scherer}, \citenamefont {Johnson}, \citenamefont {Wertheim}, \citenamefont {Otten}, \citenamefont {Aadit}, \citenamefont {Alexeev}, \citenamefont {Bresniker}, \citenamefont {Camsari}, \citenamefont {Chapman}, \citenamefont {Chatterjee}, \citenamefont {Dagnew}, \citenamefont {Esposito}, \citenamefont {Fahim}, \citenamefont {Fiorentino}, \citenamefont {Gajjar}, \citenamefont {Khalid}, \citenamefont {Kong}, \citenamefont {Kulchytskyy}, \citenamefont {Kyoseva}, \citenamefont {Li}, \citenamefont {Lott}, \citenamefont {Markov}, \citenamefont {McDermott}, \citenamefont {Pedretti}, \citenamefont {Rao}, \citenamefont {Rieffel}, \citenamefont {Silva}, \citenamefont {Sorebo}, \citenamefont {Spentzouris}, \citenamefont {Steiner}, \citenamefont {Torosov}, \citenamefont {Venturelli}, \citenamefont {Visser}, \citenamefont {Webb}, \citenamefont {Zhan}, \citenamefont {Cohen}, \citenamefont {Ronagh}, \citenamefont {Ho},
  \citenamefont {Beausoleil},\ and\ \citenamefont {Martinis}}]{mohseni2025buildquantumsupercomputerscaling}%
  \BibitemOpen
  \bibfield  {author} {\bibinfo {author} {\bibfnamefont {M.}~\bibnamefont {Mohseni}}, \bibinfo {author} {\bibfnamefont {A.}~\bibnamefont {Scherer}}, \bibinfo {author} {\bibfnamefont {K.~G.}\ \bibnamefont {Johnson}}, \bibinfo {author} {\bibfnamefont {O.}~\bibnamefont {Wertheim}}, \bibinfo {author} {\bibfnamefont {M.}~\bibnamefont {Otten}}, \bibinfo {author} {\bibfnamefont {N.~A.}\ \bibnamefont {Aadit}}, \bibinfo {author} {\bibfnamefont {Y.}~\bibnamefont {Alexeev}}, \bibinfo {author} {\bibfnamefont {K.~M.}\ \bibnamefont {Bresniker}}, \bibinfo {author} {\bibfnamefont {K.~Y.}\ \bibnamefont {Camsari}}, \bibinfo {author} {\bibfnamefont {B.}~\bibnamefont {Chapman}}, \bibinfo {author} {\bibfnamefont {S.}~\bibnamefont {Chatterjee}}, \bibinfo {author} {\bibfnamefont {G.~A.}\ \bibnamefont {Dagnew}}, \bibinfo {author} {\bibfnamefont {A.}~\bibnamefont {Esposito}}, \bibinfo {author} {\bibfnamefont {F.}~\bibnamefont {Fahim}}, \bibinfo {author} {\bibfnamefont {M.}~\bibnamefont {Fiorentino}}, \bibinfo {author} {\bibfnamefont
  {A.}~\bibnamefont {Gajjar}}, \bibinfo {author} {\bibfnamefont {A.}~\bibnamefont {Khalid}}, \bibinfo {author} {\bibfnamefont {X.}~\bibnamefont {Kong}}, \bibinfo {author} {\bibfnamefont {B.}~\bibnamefont {Kulchytskyy}}, \bibinfo {author} {\bibfnamefont {E.}~\bibnamefont {Kyoseva}}, \bibinfo {author} {\bibfnamefont {R.}~\bibnamefont {Li}}, \bibinfo {author} {\bibfnamefont {P.~A.}\ \bibnamefont {Lott}}, \bibinfo {author} {\bibfnamefont {I.~L.}\ \bibnamefont {Markov}}, \bibinfo {author} {\bibfnamefont {R.~F.}\ \bibnamefont {McDermott}}, \bibinfo {author} {\bibfnamefont {G.}~\bibnamefont {Pedretti}}, \bibinfo {author} {\bibfnamefont {P.}~\bibnamefont {Rao}}, \bibinfo {author} {\bibfnamefont {E.}~\bibnamefont {Rieffel}}, \bibinfo {author} {\bibfnamefont {A.}~\bibnamefont {Silva}}, \bibinfo {author} {\bibfnamefont {J.}~\bibnamefont {Sorebo}}, \bibinfo {author} {\bibfnamefont {P.}~\bibnamefont {Spentzouris}}, \bibinfo {author} {\bibfnamefont {Z.}~\bibnamefont {Steiner}}, \bibinfo {author} {\bibfnamefont
  {B.}~\bibnamefont {Torosov}}, \bibinfo {author} {\bibfnamefont {D.}~\bibnamefont {Venturelli}}, \bibinfo {author} {\bibfnamefont {R.~J.}\ \bibnamefont {Visser}}, \bibinfo {author} {\bibfnamefont {Z.}~\bibnamefont {Webb}}, \bibinfo {author} {\bibfnamefont {X.}~\bibnamefont {Zhan}}, \bibinfo {author} {\bibfnamefont {Y.}~\bibnamefont {Cohen}}, \bibinfo {author} {\bibfnamefont {P.}~\bibnamefont {Ronagh}}, \bibinfo {author} {\bibfnamefont {A.}~\bibnamefont {Ho}}, \bibinfo {author} {\bibfnamefont {R.~G.}\ \bibnamefont {Beausoleil}},\ and\ \bibinfo {author} {\bibfnamefont {J.~M.}\ \bibnamefont {Martinis}},\ }\href {https://arxiv.org/abs/2411.10406} {\bibfield  {journal} {\bibinfo  {journal} {arXiv:2411.10406}\ } (\bibinfo {year} {2025})}\BibitemShut {NoStop}%
\bibitem [{\citenamefont {Litinski}(2023)}]{litinski2023compute256bitellipticcurve}%
  \BibitemOpen
  \bibfield  {author} {\bibinfo {author} {\bibfnamefont {D.}~\bibnamefont {Litinski}},\ }\href {https://arxiv.org/abs/2306.08585} {\bibfield  {journal} {\bibinfo  {journal} {arXiv:2306.08585}\ } (\bibinfo {year} {2023})}\BibitemShut {NoStop}%
\bibitem [{\citenamefont {{Photonic Inc.}}(2024)}]{inc2024distributedquantumcomputingsilicon}%
  \BibitemOpen
  \bibfield  {author} {\bibinfo {author} {\bibnamefont {{Photonic Inc.}}},\ }\href {https://arxiv.org/abs/2406.01704} {\bibfield  {journal} {\bibinfo  {journal} {arXiv:2406.01704}\ } (\bibinfo {year} {2024})}\BibitemShut {NoStop}%
\bibitem [{\citenamefont {de~Gliniasty}\ \emph {et~al.}(2024)\citenamefont {de~Gliniasty}, \citenamefont {Hilaire}, \citenamefont {Emeriau}, \citenamefont {Wein}, \citenamefont {Salavrakos},\ and\ \citenamefont {Mansfield}}]{deGliniasty2024spinopticalquantum}%
  \BibitemOpen
  \bibfield  {author} {\bibinfo {author} {\bibfnamefont {G.}~\bibnamefont {de~Gliniasty}}, \bibinfo {author} {\bibfnamefont {P.}~\bibnamefont {Hilaire}}, \bibinfo {author} {\bibfnamefont {P.-E.}\ \bibnamefont {Emeriau}}, \bibinfo {author} {\bibfnamefont {S.~C.}\ \bibnamefont {Wein}}, \bibinfo {author} {\bibfnamefont {A.}~\bibnamefont {Salavrakos}},\ and\ \bibinfo {author} {\bibfnamefont {S.}~\bibnamefont {Mansfield}},\ }\href {https://doi.org/10.22331/q-2024-07-24-1423} {\bibfield  {journal} {\bibinfo  {journal} {{Quantum}}\ }\textbf {\bibinfo {volume} {8}},\ \bibinfo {pages} {1423} (\bibinfo {year} {2024})}\BibitemShut {NoStop}%
\bibitem [{\citenamefont {Riedel}\ \emph {et~al.}(2025)\citenamefont {Riedel}, \citenamefont {Graziosi}, \citenamefont {Wang}, \citenamefont {De-Eknamkul}, \citenamefont {Abulnaga}, \citenamefont {Dietz}, \citenamefont {Mucchietto}, \citenamefont {Haas}, \citenamefont {Sutula}, \citenamefont {Barral}, \citenamefont {Pompili}, \citenamefont {Robens}, \citenamefont {Ha}, \citenamefont {Sukachev}, \citenamefont {Levonian}, \citenamefont {Bhaskar}, \citenamefont {Markham},\ and\ \citenamefont {Machielse}}]{riedel2025scalablephotonicquantuminterconnect}%
  \BibitemOpen
  \bibfield  {author} {\bibinfo {author} {\bibfnamefont {D.}~\bibnamefont {Riedel}}, \bibinfo {author} {\bibfnamefont {T.}~\bibnamefont {Graziosi}}, \bibinfo {author} {\bibfnamefont {Z.}~\bibnamefont {Wang}}, \bibinfo {author} {\bibfnamefont {C.}~\bibnamefont {De-Eknamkul}}, \bibinfo {author} {\bibfnamefont {A.}~\bibnamefont {Abulnaga}}, \bibinfo {author} {\bibfnamefont {J.}~\bibnamefont {Dietz}}, \bibinfo {author} {\bibfnamefont {A.}~\bibnamefont {Mucchietto}}, \bibinfo {author} {\bibfnamefont {M.}~\bibnamefont {Haas}}, \bibinfo {author} {\bibfnamefont {M.}~\bibnamefont {Sutula}}, \bibinfo {author} {\bibfnamefont {P.}~\bibnamefont {Barral}}, \bibinfo {author} {\bibfnamefont {M.}~\bibnamefont {Pompili}}, \bibinfo {author} {\bibfnamefont {C.}~\bibnamefont {Robens}}, \bibinfo {author} {\bibfnamefont {J.}~\bibnamefont {Ha}}, \bibinfo {author} {\bibfnamefont {D.}~\bibnamefont {Sukachev}}, \bibinfo {author} {\bibfnamefont {D.}~\bibnamefont {Levonian}}, \bibinfo {author} {\bibfnamefont {M.}~\bibnamefont {Bhaskar}},
  \bibinfo {author} {\bibfnamefont {M.}~\bibnamefont {Markham}},\ and\ \bibinfo {author} {\bibfnamefont {B.}~\bibnamefont {Machielse}},\ }\href {https://arxiv.org/abs/2508.06675} {\bibfield  {journal} {\bibinfo  {journal} {arXiv:2508.06675}\ } (\bibinfo {year} {2025})}\BibitemShut {NoStop}%
\bibitem [{\citenamefont {Weaver}\ \emph {et~al.}(2024)\citenamefont {Weaver}, \citenamefont {Duivestein}, \citenamefont {Bernasconi}, \citenamefont {Scharmer}, \citenamefont {Lemang}, \citenamefont {Thiel}, \citenamefont {Hijazi}, \citenamefont {Hensen}, \citenamefont {Gr{\"o}blacher},\ and\ \citenamefont {Stockill}}]{weaver2024integrated}%
  \BibitemOpen
  \bibfield  {author} {\bibinfo {author} {\bibfnamefont {M.~J.}\ \bibnamefont {Weaver}}, \bibinfo {author} {\bibfnamefont {P.}~\bibnamefont {Duivestein}}, \bibinfo {author} {\bibfnamefont {A.~C.}\ \bibnamefont {Bernasconi}}, \bibinfo {author} {\bibfnamefont {S.}~\bibnamefont {Scharmer}}, \bibinfo {author} {\bibfnamefont {M.}~\bibnamefont {Lemang}}, \bibinfo {author} {\bibfnamefont {T.~C.~v.}\ \bibnamefont {Thiel}}, \bibinfo {author} {\bibfnamefont {F.}~\bibnamefont {Hijazi}}, \bibinfo {author} {\bibfnamefont {B.}~\bibnamefont {Hensen}}, \bibinfo {author} {\bibfnamefont {S.}~\bibnamefont {Gr{\"o}blacher}},\ and\ \bibinfo {author} {\bibfnamefont {R.}~\bibnamefont {Stockill}},\ }\href {https://www.nature.com/articles/s41565-023-01515-y} {\bibfield  {journal} {\bibinfo  {journal} {Nature nanotechnology}\ }\textbf {\bibinfo {volume} {19}},\ \bibinfo {pages} {166} (\bibinfo {year} {2024})}\BibitemShut {NoStop}%
\bibitem [{\citenamefont {Slysz}\ \emph {et~al.}(2025)\citenamefont {Slysz}, \citenamefont {Rydlichowski}, \citenamefont {Kurowski}, \citenamefont {Bacarezza}, \citenamefont {Gomez}, \citenamefont {Chandani}, \citenamefont {Heim}, \citenamefont {Khalate}, \citenamefont {Clements},\ and\ \citenamefont {Fletcher}}]{slysz2025hybrid}%
  \BibitemOpen
  \bibfield  {author} {\bibinfo {author} {\bibfnamefont {M.}~\bibnamefont {Slysz}}, \bibinfo {author} {\bibfnamefont {P.}~\bibnamefont {Rydlichowski}}, \bibinfo {author} {\bibfnamefont {K.}~\bibnamefont {Kurowski}}, \bibinfo {author} {\bibfnamefont {O.}~\bibnamefont {Bacarezza}}, \bibinfo {author} {\bibfnamefont {E.~C.}\ \bibnamefont {Gomez}}, \bibinfo {author} {\bibfnamefont {Z.}~\bibnamefont {Chandani}}, \bibinfo {author} {\bibfnamefont {B.}~\bibnamefont {Heim}}, \bibinfo {author} {\bibfnamefont {P.}~\bibnamefont {Khalate}}, \bibinfo {author} {\bibfnamefont {W.~R.}\ \bibnamefont {Clements}},\ and\ \bibinfo {author} {\bibfnamefont {J.}~\bibnamefont {Fletcher}},\ }\href {https://arxiv.org/abs/2508.16297} {\bibfield  {journal} {\bibinfo  {journal} {arXiv:2508.16297}\ } (\bibinfo {year} {2025})}\BibitemShut {NoStop}%
\bibitem [{\citenamefont {Rad}\ \emph {et~al.}(2025)\citenamefont {Rad}, \citenamefont {Ainsworth}, \citenamefont {Alexander}, \citenamefont {Altieri}, \citenamefont {Askarani}, \citenamefont {Baby}, \citenamefont {Banchi}, \citenamefont {Baragiola}, \citenamefont {Bourassa}, \citenamefont {Chadwick} \emph {et~al.}}]{AghaeeRad2025}%
  \BibitemOpen
  \bibfield  {author} {\bibinfo {author} {\bibfnamefont {H.~A.}\ \bibnamefont {Rad}}, \bibinfo {author} {\bibfnamefont {T.}~\bibnamefont {Ainsworth}}, \bibinfo {author} {\bibfnamefont {R.~N.}\ \bibnamefont {Alexander}}, \bibinfo {author} {\bibfnamefont {B.}~\bibnamefont {Altieri}}, \bibinfo {author} {\bibfnamefont {M.~F.}\ \bibnamefont {Askarani}}, \bibinfo {author} {\bibfnamefont {R.}~\bibnamefont {Baby}}, \bibinfo {author} {\bibfnamefont {L.}~\bibnamefont {Banchi}}, \bibinfo {author} {\bibfnamefont {B.~Q.}\ \bibnamefont {Baragiola}}, \bibinfo {author} {\bibfnamefont {J.~E.}\ \bibnamefont {Bourassa}}, \bibinfo {author} {\bibfnamefont {R.~S.}\ \bibnamefont {Chadwick}}, \emph {et~al.},\ }\href {https://doi.org/10.1038/s41586-024-08406-9} {\bibfield  {journal} {\bibinfo  {journal} {Nature}\ }\textbf {\bibinfo {volume} {638}},\ \bibinfo {pages} {912} (\bibinfo {year} {2025})}\BibitemShut {NoStop}%
\bibitem [{\citenamefont {Kailasanathan}\ \emph {et~al.}(2025)\citenamefont {Kailasanathan}, \citenamefont {Clements}, \citenamefont {Boskabadi}, \citenamefont {Gibford}, \citenamefont {Papadakis}, \citenamefont {Savoie},\ and\ \citenamefont {Mansouri}}]{kailasanathan2025quantumenhancedensemblegans}%
  \BibitemOpen
  \bibfield  {author} {\bibinfo {author} {\bibfnamefont {R.}~\bibnamefont {Kailasanathan}}, \bibinfo {author} {\bibfnamefont {W.~R.}\ \bibnamefont {Clements}}, \bibinfo {author} {\bibfnamefont {M.~R.}\ \bibnamefont {Boskabadi}}, \bibinfo {author} {\bibfnamefont {S.~M.}\ \bibnamefont {Gibford}}, \bibinfo {author} {\bibfnamefont {E.}~\bibnamefont {Papadakis}}, \bibinfo {author} {\bibfnamefont {C.~J.}\ \bibnamefont {Savoie}},\ and\ \bibinfo {author} {\bibfnamefont {S.~S.}\ \bibnamefont {Mansouri}},\ }\href {https://arxiv.org/abs/2508.21438} {\bibfield  {journal} {\bibinfo  {journal} {arXiv:2508.21438}\ } (\bibinfo {year} {2025})}\BibitemShut {NoStop}%
\bibitem [{\citenamefont {Larsen}\ \emph {et~al.}(2025)\citenamefont {Larsen}, \citenamefont {Bourassa}, \citenamefont {Kocsis}, \citenamefont {Tasker}, \citenamefont {Chadwick}, \citenamefont {Gonz{\'a}lez-Arciniegas}, \citenamefont {Hastrup}, \citenamefont {Lopetegui-Gonz{\'a}lez}, \citenamefont {Miatto}, \citenamefont {Motamedi} \emph {et~al.}}]{larsen2025integrated}%
  \BibitemOpen
  \bibfield  {author} {\bibinfo {author} {\bibfnamefont {M.}~\bibnamefont {Larsen}}, \bibinfo {author} {\bibfnamefont {J.}~\bibnamefont {Bourassa}}, \bibinfo {author} {\bibfnamefont {S.}~\bibnamefont {Kocsis}}, \bibinfo {author} {\bibfnamefont {J.}~\bibnamefont {Tasker}}, \bibinfo {author} {\bibfnamefont {R.}~\bibnamefont {Chadwick}}, \bibinfo {author} {\bibfnamefont {C.}~\bibnamefont {Gonz{\'a}lez-Arciniegas}}, \bibinfo {author} {\bibfnamefont {J.}~\bibnamefont {Hastrup}}, \bibinfo {author} {\bibfnamefont {C.}~\bibnamefont {Lopetegui-Gonz{\'a}lez}}, \bibinfo {author} {\bibfnamefont {F.}~\bibnamefont {Miatto}}, \bibinfo {author} {\bibfnamefont {A.}~\bibnamefont {Motamedi}}, \emph {et~al.},\ }\href {https://www.nature.com/articles/s41586-025-09044-5} {\bibfield  {journal} {\bibinfo  {journal} {Nature}\ ,\ \bibinfo {pages} {1}} (\bibinfo {year} {2025})}\BibitemShut {NoStop}%
\bibitem [{\citenamefont {Bartolucci}\ \emph {et~al.}(2021{\natexlab{a}})\citenamefont {Bartolucci}, \citenamefont {Birchall}, \citenamefont {Gimeno-Segovia}, \citenamefont {Johnston}, \citenamefont {Kieling}, \citenamefont {Pant}, \citenamefont {Rudolph}, \citenamefont {Smith}, \citenamefont {Sparrow},\ and\ \citenamefont {Vidrighin}}]{bartolucci2021creationentangledphotonicstates}%
  \BibitemOpen
  \bibfield  {author} {\bibinfo {author} {\bibfnamefont {S.}~\bibnamefont {Bartolucci}}, \bibinfo {author} {\bibfnamefont {P.~M.}\ \bibnamefont {Birchall}}, \bibinfo {author} {\bibfnamefont {M.}~\bibnamefont {Gimeno-Segovia}}, \bibinfo {author} {\bibfnamefont {E.}~\bibnamefont {Johnston}}, \bibinfo {author} {\bibfnamefont {K.}~\bibnamefont {Kieling}}, \bibinfo {author} {\bibfnamefont {M.}~\bibnamefont {Pant}}, \bibinfo {author} {\bibfnamefont {T.}~\bibnamefont {Rudolph}}, \bibinfo {author} {\bibfnamefont {J.}~\bibnamefont {Smith}}, \bibinfo {author} {\bibfnamefont {C.}~\bibnamefont {Sparrow}},\ and\ \bibinfo {author} {\bibfnamefont {M.~D.}\ \bibnamefont {Vidrighin}},\ }\href {https://arxiv.org/abs/2106.13825} {\bibfield  {journal} {\bibinfo  {journal} {arXiv:2106.13825}\ } (\bibinfo {year} {2021}{\natexlab{a}})}\BibitemShut {NoStop}%
\bibitem [{\citenamefont {Bartolucci}\ \emph {et~al.}(2021{\natexlab{b}})\citenamefont {Bartolucci}, \citenamefont {Birchall}, \citenamefont {Bonneau}, \citenamefont {Cable}, \citenamefont {Gimeno-Segovia}, \citenamefont {Kieling}, \citenamefont {Nickerson}, \citenamefont {Rudolph},\ and\ \citenamefont {Sparrow}}]{bartolucci2021switchnetworksphotonicfusionbased}%
  \BibitemOpen
  \bibfield  {author} {\bibinfo {author} {\bibfnamefont {S.}~\bibnamefont {Bartolucci}}, \bibinfo {author} {\bibfnamefont {P.}~\bibnamefont {Birchall}}, \bibinfo {author} {\bibfnamefont {D.}~\bibnamefont {Bonneau}}, \bibinfo {author} {\bibfnamefont {H.}~\bibnamefont {Cable}}, \bibinfo {author} {\bibfnamefont {M.}~\bibnamefont {Gimeno-Segovia}}, \bibinfo {author} {\bibfnamefont {K.}~\bibnamefont {Kieling}}, \bibinfo {author} {\bibfnamefont {N.}~\bibnamefont {Nickerson}}, \bibinfo {author} {\bibfnamefont {T.}~\bibnamefont {Rudolph}},\ and\ \bibinfo {author} {\bibfnamefont {C.}~\bibnamefont {Sparrow}},\ }\href {https://arxiv.org/abs/2109.13760} {\bibfield  {journal} {\bibinfo  {journal} {arXiv:2109.13760}\ } (\bibinfo {year} {2021}{\natexlab{b}})}\BibitemShut {NoStop}%
\bibitem [{\citenamefont {Zektzer}\ \emph {et~al.}(2024)\citenamefont {Zektzer}, \citenamefont {Lu}, \citenamefont {Hoang}, \citenamefont {Shrestha}, \citenamefont {Austin}, \citenamefont {Zhou}, \citenamefont {Chanana}, \citenamefont {Holland}, \citenamefont {Westly}, \citenamefont {Lett}, \citenamefont {Gorshkov},\ and\ \citenamefont {Srinivasan}}]{Zektzer:24}%
  \BibitemOpen
  \bibfield  {author} {\bibinfo {author} {\bibfnamefont {R.}~\bibnamefont {Zektzer}}, \bibinfo {author} {\bibfnamefont {X.}~\bibnamefont {Lu}}, \bibinfo {author} {\bibfnamefont {K.~T.}\ \bibnamefont {Hoang}}, \bibinfo {author} {\bibfnamefont {R.}~\bibnamefont {Shrestha}}, \bibinfo {author} {\bibfnamefont {S.}~\bibnamefont {Austin}}, \bibinfo {author} {\bibfnamefont {F.}~\bibnamefont {Zhou}}, \bibinfo {author} {\bibfnamefont {A.}~\bibnamefont {Chanana}}, \bibinfo {author} {\bibfnamefont {G.}~\bibnamefont {Holland}}, \bibinfo {author} {\bibfnamefont {D.}~\bibnamefont {Westly}}, \bibinfo {author} {\bibfnamefont {P.}~\bibnamefont {Lett}}, \bibinfo {author} {\bibfnamefont {A.~V.}\ \bibnamefont {Gorshkov}},\ and\ \bibinfo {author} {\bibfnamefont {K.}~\bibnamefont {Srinivasan}},\ }\href {https://doi.org/10.1364/OPTICA.525689} {\bibfield  {journal} {\bibinfo  {journal} {Optica}\ }\textbf {\bibinfo {volume} {11}},\ \bibinfo {pages} {1376} (\bibinfo {year} {2024})}\BibitemShut {NoStop}%
\bibitem [{\citenamefont {Bechler}\ \emph {et~al.}(2018)\citenamefont {Bechler}, \citenamefont {Borne}, \citenamefont {Rosenblum}, \citenamefont {Guendelman}, \citenamefont {Mor}, \citenamefont {Netser}, \citenamefont {Ohana}, \citenamefont {Aqua}, \citenamefont {Drucker}, \citenamefont {Finkelstein} \emph {et~al.}}]{bechler2018passive}%
  \BibitemOpen
  \bibfield  {author} {\bibinfo {author} {\bibfnamefont {O.}~\bibnamefont {Bechler}}, \bibinfo {author} {\bibfnamefont {A.}~\bibnamefont {Borne}}, \bibinfo {author} {\bibfnamefont {S.}~\bibnamefont {Rosenblum}}, \bibinfo {author} {\bibfnamefont {G.}~\bibnamefont {Guendelman}}, \bibinfo {author} {\bibfnamefont {O.~E.}\ \bibnamefont {Mor}}, \bibinfo {author} {\bibfnamefont {M.}~\bibnamefont {Netser}}, \bibinfo {author} {\bibfnamefont {T.}~\bibnamefont {Ohana}}, \bibinfo {author} {\bibfnamefont {Z.}~\bibnamefont {Aqua}}, \bibinfo {author} {\bibfnamefont {N.}~\bibnamefont {Drucker}}, \bibinfo {author} {\bibfnamefont {R.}~\bibnamefont {Finkelstein}}, \emph {et~al.},\ }\href {https://www.nature.com/articles/s41567-018-0241-6} {\bibfield  {journal} {\bibinfo  {journal} {Nature Physics}\ }\textbf {\bibinfo {volume} {14}},\ \bibinfo {pages} {996} (\bibinfo {year} {2018})}\BibitemShut {NoStop}%
\bibitem [{\citenamefont {Simmons}(2024)}]{PRXQuantum.5.010102}%
  \BibitemOpen
  \bibfield  {author} {\bibinfo {author} {\bibfnamefont {S.}~\bibnamefont {Simmons}},\ }\href {https://doi.org/10.1103/PRXQuantum.5.010102} {\bibfield  {journal} {\bibinfo  {journal} {PRX Quantum}\ }\textbf {\bibinfo {volume} {5}},\ \bibinfo {pages} {010102} (\bibinfo {year} {2024})}\BibitemShut {NoStop}%
\bibitem [{\citenamefont {Chan}\ \emph {et~al.}(2025)\citenamefont {Chan}, \citenamefont {Capatos}, \citenamefont {Lodahl}, \citenamefont {Sørensen},\ and\ \citenamefont {Paesani}}]{chan2025practicalblueprint}%
  \BibitemOpen
  \bibfield  {author} {\bibinfo {author} {\bibfnamefont {M.~L.}\ \bibnamefont {Chan}}, \bibinfo {author} {\bibfnamefont {A.~A.}\ \bibnamefont {Capatos}}, \bibinfo {author} {\bibfnamefont {P.}~\bibnamefont {Lodahl}}, \bibinfo {author} {\bibfnamefont {A.~S.}\ \bibnamefont {Sørensen}},\ and\ \bibinfo {author} {\bibfnamefont {S.}~\bibnamefont {Paesani}},\ }\href {https://arxiv.org/abs/2507.16152} {\bibfield  {journal} {\bibinfo  {journal} {arXiv:2507.16152}\ } (\bibinfo {year} {2025})}\BibitemShut {NoStop}%
\bibitem [{\citenamefont {Dessertaine}\ \emph {et~al.}(2026)\citenamefont {Dessertaine}, \citenamefont {Bourdoncle}, \citenamefont {Denys}, \citenamefont {de~Gliniasty}, \citenamefont {d'Istria}, \citenamefont {Valentí-Rojas}, \citenamefont {Mansfield},\ and\ \citenamefont {Hilaire}}]{dessertaine2026enhancedfaulttolerancephotonicquantum}%
  \BibitemOpen
  \bibfield  {author} {\bibinfo {author} {\bibfnamefont {T.}~\bibnamefont {Dessertaine}}, \bibinfo {author} {\bibfnamefont {B.}~\bibnamefont {Bourdoncle}}, \bibinfo {author} {\bibfnamefont {A.}~\bibnamefont {Denys}}, \bibinfo {author} {\bibfnamefont {G.}~\bibnamefont {de~Gliniasty}}, \bibinfo {author} {\bibfnamefont {P.~C.}\ \bibnamefont {d'Istria}}, \bibinfo {author} {\bibfnamefont {G.}~\bibnamefont {Valentí-Rojas}}, \bibinfo {author} {\bibfnamefont {S.}~\bibnamefont {Mansfield}},\ and\ \bibinfo {author} {\bibfnamefont {P.}~\bibnamefont {Hilaire}},\ }\href {https://arxiv.org/abs/2410.07065} {\bibfield  {journal} {\bibinfo  {journal} {arXiv}\ } (\bibinfo {year} {2026})},\ \Eprint {https://arxiv.org/abs/2410.07065} {2410.07065} \BibitemShut {NoStop}%
\bibitem [{\citenamefont {Canteri}\ \emph {et~al.}(2025)\citenamefont {Canteri}, \citenamefont {Koong}, \citenamefont {Bate}, \citenamefont {Winkler}, \citenamefont {Krutyanskiy},\ and\ \citenamefont {Lanyon}}]{v5k1-whwz}%
  \BibitemOpen
  \bibfield  {author} {\bibinfo {author} {\bibfnamefont {M.}~\bibnamefont {Canteri}}, \bibinfo {author} {\bibfnamefont {Z.~X.}\ \bibnamefont {Koong}}, \bibinfo {author} {\bibfnamefont {J.}~\bibnamefont {Bate}}, \bibinfo {author} {\bibfnamefont {A.}~\bibnamefont {Winkler}}, \bibinfo {author} {\bibfnamefont {V.}~\bibnamefont {Krutyanskiy}},\ and\ \bibinfo {author} {\bibfnamefont {B.~P.}\ \bibnamefont {Lanyon}},\ }\href {https://doi.org/10.1103/v5k1-whwz} {\bibfield  {journal} {\bibinfo  {journal} {Physical Review Letters}\ }\textbf {\bibinfo {volume} {135}},\ \bibinfo {pages} {080801} (\bibinfo {year} {2025})}\BibitemShut {NoStop}%
\bibitem [{\citenamefont {Stephenson}\ \emph {et~al.}(2020)\citenamefont {Stephenson}, \citenamefont {Nadlinger}, \citenamefont {Nichol}, \citenamefont {An}, \citenamefont {Drmota}, \citenamefont {Ballance}, \citenamefont {Thirumalai}, \citenamefont {Goodwin}, \citenamefont {Lucas},\ and\ \citenamefont {Ballance}}]{stephenson2020high}%
  \BibitemOpen
  \bibfield  {author} {\bibinfo {author} {\bibfnamefont {L.}~\bibnamefont {Stephenson}}, \bibinfo {author} {\bibfnamefont {D.}~\bibnamefont {Nadlinger}}, \bibinfo {author} {\bibfnamefont {B.}~\bibnamefont {Nichol}}, \bibinfo {author} {\bibfnamefont {S.}~\bibnamefont {An}}, \bibinfo {author} {\bibfnamefont {P.}~\bibnamefont {Drmota}}, \bibinfo {author} {\bibfnamefont {T.~G.}\ \bibnamefont {Ballance}}, \bibinfo {author} {\bibfnamefont {K.}~\bibnamefont {Thirumalai}}, \bibinfo {author} {\bibfnamefont {J.~F.}\ \bibnamefont {Goodwin}}, \bibinfo {author} {\bibfnamefont {D.~M.}\ \bibnamefont {Lucas}},\ and\ \bibinfo {author} {\bibfnamefont {C.}~\bibnamefont {Ballance}},\ }\href {https://journals.aps.org/prl/abstract/10.1103/PhysRevLett.124.110501} {\bibfield  {journal} {\bibinfo  {journal} {Physical Review Letters}\ }\textbf {\bibinfo {volume} {124}},\ \bibinfo {pages} {110501} (\bibinfo {year} {2020})}\BibitemShut {NoStop}%
\bibitem [{\citenamefont {Krutyanskiy}\ \emph {et~al.}(2023)\citenamefont {Krutyanskiy}, \citenamefont {Galli}, \citenamefont {Krcmarsky}, \citenamefont {Baier}, \citenamefont {Fioretto}, \citenamefont {Pu}, \citenamefont {Mazloom}, \citenamefont {Sekatski}, \citenamefont {Canteri}, \citenamefont {Teller} \emph {et~al.}}]{krutyanskiy2023entanglement}%
  \BibitemOpen
  \bibfield  {author} {\bibinfo {author} {\bibfnamefont {V.}~\bibnamefont {Krutyanskiy}}, \bibinfo {author} {\bibfnamefont {M.}~\bibnamefont {Galli}}, \bibinfo {author} {\bibfnamefont {V.}~\bibnamefont {Krcmarsky}}, \bibinfo {author} {\bibfnamefont {S.}~\bibnamefont {Baier}}, \bibinfo {author} {\bibfnamefont {D.}~\bibnamefont {Fioretto}}, \bibinfo {author} {\bibfnamefont {Y.}~\bibnamefont {Pu}}, \bibinfo {author} {\bibfnamefont {A.}~\bibnamefont {Mazloom}}, \bibinfo {author} {\bibfnamefont {P.}~\bibnamefont {Sekatski}}, \bibinfo {author} {\bibfnamefont {M.}~\bibnamefont {Canteri}}, \bibinfo {author} {\bibfnamefont {M.}~\bibnamefont {Teller}}, \emph {et~al.},\ }\href {https://journals.aps.org/prl/abstract/10.1103/PhysRevLett.130.050803} {\bibfield  {journal} {\bibinfo  {journal} {Physical Review Letters}\ }\textbf {\bibinfo {volume} {130}},\ \bibinfo {pages} {050803} (\bibinfo {year} {2023})}\BibitemShut {NoStop}%
\bibitem [{\citenamefont {DeAbreu}\ \emph {et~al.}(2023)\citenamefont {DeAbreu}, \citenamefont {Bowness}, \citenamefont {Alizadeh}, \citenamefont {Chartrand}, \citenamefont {Brunelle}, \citenamefont {MacQuarrie}, \citenamefont {Lee-Hone}, \citenamefont {Ruether}, \citenamefont {Kazemi}, \citenamefont {Kurkjian} \emph {et~al.}}]{deabreu2023waveguide}%
  \BibitemOpen
  \bibfield  {author} {\bibinfo {author} {\bibfnamefont {A.}~\bibnamefont {DeAbreu}}, \bibinfo {author} {\bibfnamefont {C.}~\bibnamefont {Bowness}}, \bibinfo {author} {\bibfnamefont {A.}~\bibnamefont {Alizadeh}}, \bibinfo {author} {\bibfnamefont {C.}~\bibnamefont {Chartrand}}, \bibinfo {author} {\bibfnamefont {N.}~\bibnamefont {Brunelle}}, \bibinfo {author} {\bibfnamefont {E.}~\bibnamefont {MacQuarrie}}, \bibinfo {author} {\bibfnamefont {N.}~\bibnamefont {Lee-Hone}}, \bibinfo {author} {\bibfnamefont {M.}~\bibnamefont {Ruether}}, \bibinfo {author} {\bibfnamefont {M.}~\bibnamefont {Kazemi}}, \bibinfo {author} {\bibfnamefont {A.}~\bibnamefont {Kurkjian}}, \emph {et~al.},\ }\href {https://opg.optica.org/oe/fulltext.cfm?uri=oe-31-9-15045} {\bibfield  {journal} {\bibinfo  {journal} {Optics Express}\ }\textbf {\bibinfo {volume} {31}},\ \bibinfo {pages} {15045} (\bibinfo {year} {2023})}\BibitemShut {NoStop}%
\bibitem [{\citenamefont {Lodahl}\ \emph {et~al.}(2015)\citenamefont {Lodahl}, \citenamefont {Mahmoodian},\ and\ \citenamefont {Stobbe}}]{lodahl2015interfacing}%
  \BibitemOpen
  \bibfield  {author} {\bibinfo {author} {\bibfnamefont {P.}~\bibnamefont {Lodahl}}, \bibinfo {author} {\bibfnamefont {S.}~\bibnamefont {Mahmoodian}},\ and\ \bibinfo {author} {\bibfnamefont {S.}~\bibnamefont {Stobbe}},\ }\href {https://journals.aps.org/rmp/abstract/10.1103/RevModPhys.87.347} {\bibfield  {journal} {\bibinfo  {journal} {Reviews of Modern Physics}\ }\textbf {\bibinfo {volume} {87}},\ \bibinfo {pages} {347} (\bibinfo {year} {2015})}\BibitemShut {NoStop}%
\bibitem [{\citenamefont {Briegel}\ \emph {et~al.}(2009)\citenamefont {Briegel}, \citenamefont {Browne}, \citenamefont {D{\"u}r}, \citenamefont {Raussendorf},\ and\ \citenamefont {Van~den Nest}}]{Briegel2009}%
  \BibitemOpen
  \bibfield  {author} {\bibinfo {author} {\bibfnamefont {H.~J.}\ \bibnamefont {Briegel}}, \bibinfo {author} {\bibfnamefont {D.~E.}\ \bibnamefont {Browne}}, \bibinfo {author} {\bibfnamefont {W.}~\bibnamefont {D{\"u}r}}, \bibinfo {author} {\bibfnamefont {R.}~\bibnamefont {Raussendorf}},\ and\ \bibinfo {author} {\bibfnamefont {M.}~\bibnamefont {Van~den Nest}},\ }\href {https://doi.org/10.1038/nphys1157} {\bibfield  {journal} {\bibinfo  {journal} {Nature Physics}\ }\textbf {\bibinfo {volume} {5}},\ \bibinfo {pages} {19} (\bibinfo {year} {2009})}\BibitemShut {NoStop}%
\bibitem [{\citenamefont {Raussendorf}\ and\ \citenamefont {Briegel}(2001)}]{PhysRevLett.86.5188}%
  \BibitemOpen
  \bibfield  {author} {\bibinfo {author} {\bibfnamefont {R.}~\bibnamefont {Raussendorf}}\ and\ \bibinfo {author} {\bibfnamefont {H.~J.}\ \bibnamefont {Briegel}},\ }\href {https://doi.org/10.1103/PhysRevLett.86.5188} {\bibfield  {journal} {\bibinfo  {journal} {Physical Review Letters}\ }\textbf {\bibinfo {volume} {86}},\ \bibinfo {pages} {5188} (\bibinfo {year} {2001})}\BibitemShut {NoStop}%
\bibitem [{\citenamefont {Bartolucci}\ \emph {et~al.}(2023)\citenamefont {Bartolucci}, \citenamefont {Birchall}, \citenamefont {Bomb{\'\i}n}, \citenamefont {Cable}, \citenamefont {Dawson}, \citenamefont {Gimeno-Segovia}, \citenamefont {Johnston}, \citenamefont {Kieling}, \citenamefont {Nickerson}, \citenamefont {Pant}, \citenamefont {Pastawski}, \citenamefont {Rudolph},\ and\ \citenamefont {Sparrow}}]{Bartolucci2023-tu}%
  \BibitemOpen
  \bibfield  {author} {\bibinfo {author} {\bibfnamefont {S.}~\bibnamefont {Bartolucci}}, \bibinfo {author} {\bibfnamefont {P.}~\bibnamefont {Birchall}}, \bibinfo {author} {\bibfnamefont {H.}~\bibnamefont {Bomb{\'\i}n}}, \bibinfo {author} {\bibfnamefont {H.}~\bibnamefont {Cable}}, \bibinfo {author} {\bibfnamefont {C.}~\bibnamefont {Dawson}}, \bibinfo {author} {\bibfnamefont {M.}~\bibnamefont {Gimeno-Segovia}}, \bibinfo {author} {\bibfnamefont {E.}~\bibnamefont {Johnston}}, \bibinfo {author} {\bibfnamefont {K.}~\bibnamefont {Kieling}}, \bibinfo {author} {\bibfnamefont {N.}~\bibnamefont {Nickerson}}, \bibinfo {author} {\bibfnamefont {M.}~\bibnamefont {Pant}}, \bibinfo {author} {\bibfnamefont {F.}~\bibnamefont {Pastawski}}, \bibinfo {author} {\bibfnamefont {T.}~\bibnamefont {Rudolph}},\ and\ \bibinfo {author} {\bibfnamefont {C.}~\bibnamefont {Sparrow}},\ }\href {https://doi.org/10.1038/s41467-023-36493-1} {\bibfield  {journal} {\bibinfo  {journal} {Nature Communications}\ }\textbf {\bibinfo {volume} {14}},\
  \bibinfo {pages} {912} (\bibinfo {year} {2023})}\BibitemShut {NoStop}%
\bibitem [{\citenamefont {Pankovich}\ \emph {et~al.}(2024{\natexlab{a}})\citenamefont {Pankovich}, \citenamefont {Kan}, \citenamefont {Wan}, \citenamefont {Ostmann}, \citenamefont {Neville}, \citenamefont {Omkar}, \citenamefont {Sohbi},\ and\ \citenamefont {Br\'adler}}]{PhysRevLett.133.050604}%
  \BibitemOpen
  \bibfield  {author} {\bibinfo {author} {\bibfnamefont {B.}~\bibnamefont {Pankovich}}, \bibinfo {author} {\bibfnamefont {A.}~\bibnamefont {Kan}}, \bibinfo {author} {\bibfnamefont {K.~H.}\ \bibnamefont {Wan}}, \bibinfo {author} {\bibfnamefont {M.}~\bibnamefont {Ostmann}}, \bibinfo {author} {\bibfnamefont {A.}~\bibnamefont {Neville}}, \bibinfo {author} {\bibfnamefont {S.}~\bibnamefont {Omkar}}, \bibinfo {author} {\bibfnamefont {A.}~\bibnamefont {Sohbi}},\ and\ \bibinfo {author} {\bibfnamefont {K.}~\bibnamefont {Br\'adler}},\ }\href {https://doi.org/10.1103/PhysRevLett.133.050604} {\bibfield  {journal} {\bibinfo  {journal} {Physical Review Letters}\ }\textbf {\bibinfo {volume} {133}},\ \bibinfo {pages} {050604} (\bibinfo {year} {2024}{\natexlab{a}})}\BibitemShut {NoStop}%
\bibitem [{\citenamefont {Bartolucci}\ \emph {et~al.}(2025)\citenamefont {Bartolucci}, \citenamefont {Bell}, \citenamefont {Bombin}, \citenamefont {Birchall}, \citenamefont {Bulmer}, \citenamefont {Dawson}, \citenamefont {Farrelly}, \citenamefont {Gartenstein}, \citenamefont {Gimeno-Segovia}, \citenamefont {Litinski}, \citenamefont {Liu}, \citenamefont {Knegjens}, \citenamefont {Nickerson}, \citenamefont {Olivo}, \citenamefont {Pant}, \citenamefont {Patil}, \citenamefont {Roberts}, \citenamefont {Rudolph}, \citenamefont {Sparrow}, \citenamefont {Tuckett},\ and\ \citenamefont {Veitia}}]{bartolucci2025comparisonschemeshighlyloss}%
  \BibitemOpen
  \bibfield  {author} {\bibinfo {author} {\bibfnamefont {S.}~\bibnamefont {Bartolucci}}, \bibinfo {author} {\bibfnamefont {T.}~\bibnamefont {Bell}}, \bibinfo {author} {\bibfnamefont {H.}~\bibnamefont {Bombin}}, \bibinfo {author} {\bibfnamefont {P.}~\bibnamefont {Birchall}}, \bibinfo {author} {\bibfnamefont {J.}~\bibnamefont {Bulmer}}, \bibinfo {author} {\bibfnamefont {C.}~\bibnamefont {Dawson}}, \bibinfo {author} {\bibfnamefont {T.}~\bibnamefont {Farrelly}}, \bibinfo {author} {\bibfnamefont {S.}~\bibnamefont {Gartenstein}}, \bibinfo {author} {\bibfnamefont {M.}~\bibnamefont {Gimeno-Segovia}}, \bibinfo {author} {\bibfnamefont {D.}~\bibnamefont {Litinski}}, \bibinfo {author} {\bibfnamefont {Y.}~\bibnamefont {Liu}}, \bibinfo {author} {\bibfnamefont {R.}~\bibnamefont {Knegjens}}, \bibinfo {author} {\bibfnamefont {N.}~\bibnamefont {Nickerson}}, \bibinfo {author} {\bibfnamefont {A.}~\bibnamefont {Olivo}}, \bibinfo {author} {\bibfnamefont {M.}~\bibnamefont {Pant}}, \bibinfo {author} {\bibfnamefont {A.}~\bibnamefont
  {Patil}}, \bibinfo {author} {\bibfnamefont {S.}~\bibnamefont {Roberts}}, \bibinfo {author} {\bibfnamefont {T.}~\bibnamefont {Rudolph}}, \bibinfo {author} {\bibfnamefont {C.}~\bibnamefont {Sparrow}}, \bibinfo {author} {\bibfnamefont {D.}~\bibnamefont {Tuckett}},\ and\ \bibinfo {author} {\bibfnamefont {A.}~\bibnamefont {Veitia}},\ }\href {https://arxiv.org/abs/2506.11975} {\bibfield  {journal} {\bibinfo  {journal} {arXiv:2506.11975}\ } (\bibinfo {year} {2025})}\BibitemShut {NoStop}%
\bibitem [{\citenamefont {Romero}\ and\ \citenamefont {Milburn}()}]{10.1093/acrefore/9780190871994.013.84}%
  \BibitemOpen
  \bibfield  {author} {\bibinfo {author} {\bibfnamefont {J.}~\bibnamefont {Romero}}\ and\ \bibinfo {author} {\bibfnamefont {G.}~\bibnamefont {Milburn}},\ }in\ \href {https://doi.org/10.1093/acrefore/9780190871994.013.84} {\emph {\bibinfo {booktitle} {Oxford Research Encyclopedia of Physics}}}\ (\bibinfo  {publisher} {Oxford University Press})\BibitemShut {NoStop}%
\bibitem [{\citenamefont {Knill}\ \emph {et~al.}(2001)\citenamefont {Knill}, \citenamefont {Laflamme},\ and\ \citenamefont {Milburn}}]{Knill2001}%
  \BibitemOpen
  \bibfield  {author} {\bibinfo {author} {\bibfnamefont {E.}~\bibnamefont {Knill}}, \bibinfo {author} {\bibfnamefont {R.}~\bibnamefont {Laflamme}},\ and\ \bibinfo {author} {\bibfnamefont {G.~J.}\ \bibnamefont {Milburn}},\ }\href {https://doi.org/10.1038/35051009} {\bibfield  {journal} {\bibinfo  {journal} {Nature}\ }\textbf {\bibinfo {volume} {409}},\ \bibinfo {pages} {46} (\bibinfo {year} {2001})}\BibitemShut {NoStop}%
\bibitem [{\citenamefont {Thomas}\ \emph {et~al.}(2022)\citenamefont {Thomas}, \citenamefont {Ruscio}, \citenamefont {Morin},\ and\ \citenamefont {Rempe}}]{Thomas2022}%
  \BibitemOpen
  \bibfield  {author} {\bibinfo {author} {\bibfnamefont {P.}~\bibnamefont {Thomas}}, \bibinfo {author} {\bibfnamefont {L.}~\bibnamefont {Ruscio}}, \bibinfo {author} {\bibfnamefont {O.}~\bibnamefont {Morin}},\ and\ \bibinfo {author} {\bibfnamefont {G.}~\bibnamefont {Rempe}},\ }\href {https://doi.org/10.1038/s41586-022-04987-5} {\bibfield  {journal} {\bibinfo  {journal} {Nature}\ }\textbf {\bibinfo {volume} {608}},\ \bibinfo {pages} {677} (\bibinfo {year} {2022})}\BibitemShut {NoStop}%
\bibitem [{\citenamefont {Nielsen}\ \emph {et~al.}(2024)\citenamefont {Nielsen}, \citenamefont {Wang}, \citenamefont {Deacon}, \citenamefont {Sund}, \citenamefont {Liu}, \citenamefont {Scholz}, \citenamefont {Wieck}, \citenamefont {Ludwig}, \citenamefont {Midolo}, \citenamefont {Sørensen}, \citenamefont {Paesani},\ and\ \citenamefont {Lodahl}}]{nielsen2024programmablenonlinearquantumphotonic}%
  \BibitemOpen
  \bibfield  {author} {\bibinfo {author} {\bibfnamefont {K.~H.}\ \bibnamefont {Nielsen}}, \bibinfo {author} {\bibfnamefont {Y.}~\bibnamefont {Wang}}, \bibinfo {author} {\bibfnamefont {E.}~\bibnamefont {Deacon}}, \bibinfo {author} {\bibfnamefont {P.~I.}\ \bibnamefont {Sund}}, \bibinfo {author} {\bibfnamefont {Z.}~\bibnamefont {Liu}}, \bibinfo {author} {\bibfnamefont {S.}~\bibnamefont {Scholz}}, \bibinfo {author} {\bibfnamefont {A.~D.}\ \bibnamefont {Wieck}}, \bibinfo {author} {\bibfnamefont {A.}~\bibnamefont {Ludwig}}, \bibinfo {author} {\bibfnamefont {L.}~\bibnamefont {Midolo}}, \bibinfo {author} {\bibfnamefont {A.~S.}\ \bibnamefont {Sørensen}}, \bibinfo {author} {\bibfnamefont {S.}~\bibnamefont {Paesani}},\ and\ \bibinfo {author} {\bibfnamefont {P.}~\bibnamefont {Lodahl}},\ }\href {https://arxiv.org/abs/2405.17941} {\bibfield  {journal} {\bibinfo  {journal} {arXiv:2405.17941}\ } (\bibinfo {year} {2024})}\BibitemShut {NoStop}%
\bibitem [{\citenamefont {Uppu}\ \emph {et~al.}(2020)\citenamefont {Uppu}, \citenamefont {Pedersen}, \citenamefont {Wang}, \citenamefont {Olesen}, \citenamefont {Papon}, \citenamefont {Zhou}, \citenamefont {Midolo}, \citenamefont {Scholz}, \citenamefont {Wieck}, \citenamefont {Ludwig},\ and\ \citenamefont {Lodahl}}]{uppu2020}%
  \BibitemOpen
  \bibfield  {author} {\bibinfo {author} {\bibfnamefont {R.}~\bibnamefont {Uppu}}, \bibinfo {author} {\bibfnamefont {F.~T.}\ \bibnamefont {Pedersen}}, \bibinfo {author} {\bibfnamefont {Y.}~\bibnamefont {Wang}}, \bibinfo {author} {\bibfnamefont {C.~T.}\ \bibnamefont {Olesen}}, \bibinfo {author} {\bibfnamefont {C.}~\bibnamefont {Papon}}, \bibinfo {author} {\bibfnamefont {X.}~\bibnamefont {Zhou}}, \bibinfo {author} {\bibfnamefont {L.}~\bibnamefont {Midolo}}, \bibinfo {author} {\bibfnamefont {S.}~\bibnamefont {Scholz}}, \bibinfo {author} {\bibfnamefont {A.~D.}\ \bibnamefont {Wieck}}, \bibinfo {author} {\bibfnamefont {A.}~\bibnamefont {Ludwig}},\ and\ \bibinfo {author} {\bibfnamefont {P.}~\bibnamefont {Lodahl}},\ }\href {https://doi.org/10.1126/sciadv.abc8268} {\bibfield  {journal} {\bibinfo  {journal} {Science Advances}\ }\textbf {\bibinfo {volume} {6}},\ \bibinfo {pages} {eabc8268} (\bibinfo {year} {2020})}\BibitemShut {NoStop}%
\bibitem [{\citenamefont {Pick}\ \emph {et~al.}(2021)\citenamefont {Pick}, \citenamefont {Matekole}, \citenamefont {Aqua}, \citenamefont {Guendelman}, \citenamefont {Firstenberg}, \citenamefont {Dowling},\ and\ \citenamefont {Dayan}}]{PickNonlinearRouter}%
  \BibitemOpen
  \bibfield  {author} {\bibinfo {author} {\bibfnamefont {A.}~\bibnamefont {Pick}}, \bibinfo {author} {\bibfnamefont {E.}~\bibnamefont {Matekole}}, \bibinfo {author} {\bibfnamefont {Z.}~\bibnamefont {Aqua}}, \bibinfo {author} {\bibfnamefont {G.}~\bibnamefont {Guendelman}}, \bibinfo {author} {\bibfnamefont {O.}~\bibnamefont {Firstenberg}}, \bibinfo {author} {\bibfnamefont {J.}~\bibnamefont {Dowling}},\ and\ \bibinfo {author} {\bibfnamefont {B.}~\bibnamefont {Dayan}},\ }\href {https://doi.org/10.1103/PhysRevApplied.15.054054} {\bibfield  {journal} {\bibinfo  {journal} {Physical Review Applied}\ }\textbf {\bibinfo {volume} {15}},\ \bibinfo {pages} {054054} (\bibinfo {year} {2021})}\BibitemShut {NoStop}%
\bibitem [{\citenamefont {Witthaut}\ \emph {et~al.}(2012)\citenamefont {Witthaut}, \citenamefont {Lukin},\ and\ \citenamefont {Sørensen}}]{Witthaut_2012}%
  \BibitemOpen
  \bibfield  {author} {\bibinfo {author} {\bibfnamefont {D.}~\bibnamefont {Witthaut}}, \bibinfo {author} {\bibfnamefont {M.~D.}\ \bibnamefont {Lukin}},\ and\ \bibinfo {author} {\bibfnamefont {A.~S.}\ \bibnamefont {Sørensen}},\ }\href {https://doi.org/10.1209/0295-5075/97/50007} {\bibfield  {journal} {\bibinfo  {journal} {Europhysics Letters}\ }\textbf {\bibinfo {volume} {97}},\ \bibinfo {pages} {50007} (\bibinfo {year} {2012})}\BibitemShut {NoStop}%
\bibitem [{\citenamefont {Wein}\ \emph {et~al.}(2024)\citenamefont {Wein}, \citenamefont {de~Brugière}, \citenamefont {Music}, \citenamefont {Senellart}, \citenamefont {Bourdoncle},\ and\ \citenamefont {Mansfield}}]{wein2024minimizingresourceoverheadfusionbased}%
  \BibitemOpen
  \bibfield  {author} {\bibinfo {author} {\bibfnamefont {S.~C.}\ \bibnamefont {Wein}}, \bibinfo {author} {\bibfnamefont {T.~G.}\ \bibnamefont {de~Brugière}}, \bibinfo {author} {\bibfnamefont {L.}~\bibnamefont {Music}}, \bibinfo {author} {\bibfnamefont {P.}~\bibnamefont {Senellart}}, \bibinfo {author} {\bibfnamefont {B.}~\bibnamefont {Bourdoncle}},\ and\ \bibinfo {author} {\bibfnamefont {S.}~\bibnamefont {Mansfield}},\ }\href {https://arxiv.org/abs/2412.08611} {\bibfield  {journal} {\bibinfo  {journal} {arXiv:2412.08611}\ } (\bibinfo {year} {2024})}\BibitemShut {NoStop}%
\bibitem [{\citenamefont {{Storz, S., Schär, J., Kulikov, A. et al.}}(2023)}]{Storz2023}%
  \BibitemOpen
  \bibfield  {author} {\bibinfo {author} {\bibnamefont {{Storz, S., Schär, J., Kulikov, A. et al.}}},\ }\href {https://doi.org/10.1038/s41586-023-05885-0} {\bibfield  {journal} {\bibinfo  {journal} {Nature}\ }\textbf {\bibinfo {volume} {617}},\ \bibinfo {pages} {265} (\bibinfo {year} {2023})}\BibitemShut {NoStop}%
\bibitem [{\citenamefont {Bolt}\ \emph {et~al.}(2016)\citenamefont {Bolt}, \citenamefont {Duclos-Cianci}, \citenamefont {Poulin},\ and\ \citenamefont {Stace}}]{PhysRevLett.117.070501}%
  \BibitemOpen
  \bibfield  {author} {\bibinfo {author} {\bibfnamefont {A.}~\bibnamefont {Bolt}}, \bibinfo {author} {\bibfnamefont {G.}~\bibnamefont {Duclos-Cianci}}, \bibinfo {author} {\bibfnamefont {D.}~\bibnamefont {Poulin}},\ and\ \bibinfo {author} {\bibfnamefont {T.~M.}\ \bibnamefont {Stace}},\ }\href {https://doi.org/10.1103/PhysRevLett.117.070501} {\bibfield  {journal} {\bibinfo  {journal} {Physical Review Letters}\ }\textbf {\bibinfo {volume} {117}},\ \bibinfo {pages} {070501} (\bibinfo {year} {2016})}\BibitemShut {NoStop}%
\bibitem [{\citenamefont {Raussendorf}\ \emph {et~al.}(2007)\citenamefont {Raussendorf}, \citenamefont {Harrington},\ and\ \citenamefont {Goyal}}]{Raussendorf_2007}%
  \BibitemOpen
  \bibfield  {author} {\bibinfo {author} {\bibfnamefont {R.}~\bibnamefont {Raussendorf}}, \bibinfo {author} {\bibfnamefont {J.}~\bibnamefont {Harrington}},\ and\ \bibinfo {author} {\bibfnamefont {K.}~\bibnamefont {Goyal}},\ }\href {https://doi.org/10.1088/1367-2630/9/6/199} {\bibfield  {journal} {\bibinfo  {journal} {New Journal of Physics}\ }\textbf {\bibinfo {volume} {9}},\ \bibinfo {pages} {199} (\bibinfo {year} {2007})}\BibitemShut {NoStop}%
\bibitem [{\citenamefont {Bravyi}\ and\ \citenamefont {Kitaev}(2005)}]{PhysRevA.71.022316}%
  \BibitemOpen
  \bibfield  {author} {\bibinfo {author} {\bibfnamefont {S.}~\bibnamefont {Bravyi}}\ and\ \bibinfo {author} {\bibfnamefont {A.}~\bibnamefont {Kitaev}},\ }\href {https://doi.org/10.1103/PhysRevA.71.022316} {\bibfield  {journal} {\bibinfo  {journal} {Physical Review A}\ }\textbf {\bibinfo {volume} {71}},\ \bibinfo {pages} {022316} (\bibinfo {year} {2005})}\BibitemShut {NoStop}%
\bibitem [{\citenamefont {Bombin}\ \emph {et~al.}(2021)\citenamefont {Bombin}, \citenamefont {Kim}, \citenamefont {Litinski}, \citenamefont {Nickerson}, \citenamefont {Pant}, \citenamefont {Pastawski}, \citenamefont {Roberts},\ and\ \citenamefont {Rudolph}}]{bombin2021interleavingmodulararchitecturesfaulttolerant}%
  \BibitemOpen
  \bibfield  {author} {\bibinfo {author} {\bibfnamefont {H.}~\bibnamefont {Bombin}}, \bibinfo {author} {\bibfnamefont {I.~H.}\ \bibnamefont {Kim}}, \bibinfo {author} {\bibfnamefont {D.}~\bibnamefont {Litinski}}, \bibinfo {author} {\bibfnamefont {N.}~\bibnamefont {Nickerson}}, \bibinfo {author} {\bibfnamefont {M.}~\bibnamefont {Pant}}, \bibinfo {author} {\bibfnamefont {F.}~\bibnamefont {Pastawski}}, \bibinfo {author} {\bibfnamefont {S.}~\bibnamefont {Roberts}},\ and\ \bibinfo {author} {\bibfnamefont {T.}~\bibnamefont {Rudolph}},\ }\href {https://arxiv.org/abs/2103.08612} {\bibfield  {journal} {\bibinfo  {journal} {arXiv:2103.08612}\ } (\bibinfo {year} {2021})}\BibitemShut {NoStop}%
\bibitem [{\citenamefont {{PsiQuantum}}(2025)}]{alexander2024manufacturableplatformphotonicquantum}%
  \BibitemOpen
  \bibfield  {author} {\bibinfo {author} {\bibnamefont {{PsiQuantum}}},\ }\href {https://doi.org/https://doi.org/10.1038/s41586-025-08820-7} {\bibfield  {journal} {\bibinfo  {journal} {Nature}\ }\textbf {\bibinfo {volume} {641}},\ \bibinfo {pages} {876} (\bibinfo {year} {2025})}\BibitemShut {NoStop}%
\bibitem [{\citenamefont {Steinmetz}\ \emph {et~al.}(2024)\citenamefont {Steinmetz}, \citenamefont {Ostmann}, \citenamefont {Neville}, \citenamefont {Pankovich},\ and\ \citenamefont {Sohbi}}]{steinmetz2024simulating}%
  \BibitemOpen
  \bibfield  {author} {\bibinfo {author} {\bibfnamefont {J.}~\bibnamefont {Steinmetz}}, \bibinfo {author} {\bibfnamefont {M.}~\bibnamefont {Ostmann}}, \bibinfo {author} {\bibfnamefont {A.}~\bibnamefont {Neville}}, \bibinfo {author} {\bibfnamefont {B.}~\bibnamefont {Pankovich}},\ and\ \bibinfo {author} {\bibfnamefont {A.}~\bibnamefont {Sohbi}},\ }\href {https://arxiv.org/abs/2412.13330} {\bibfield  {journal} {\bibinfo  {journal} {arXiv:2412.13330}\ } (\bibinfo {year} {2024})}\BibitemShut {NoStop}%
\bibitem [{\citenamefont {Bacon}\ and\ \citenamefont {Casaccino}(2006)}]{bacon2006quantumerrorcorrectingsubsystem}%
  \BibitemOpen
  \bibfield  {author} {\bibinfo {author} {\bibfnamefont {D.}~\bibnamefont {Bacon}}\ and\ \bibinfo {author} {\bibfnamefont {A.}~\bibnamefont {Casaccino}},\ }\href {https://arxiv.org/abs/quant-ph/0610088} {\bibfield  {journal} {\bibinfo  {journal} {arXiv}\ } (\bibinfo {year} {2006})},\ \Eprint {https://arxiv.org/abs/quant-ph/0610088} {quant-ph/0610088} \BibitemShut {NoStop}%
\bibitem [{\citenamefont {Pankovich}\ \emph {et~al.}(2024{\natexlab{b}})\citenamefont {Pankovich}, \citenamefont {Neville}, \citenamefont {Kan}, \citenamefont {Omkar}, \citenamefont {Wan},\ and\ \citenamefont {Br\'adler}}]{PhysRevA.110.032402}%
  \BibitemOpen
  \bibfield  {author} {\bibinfo {author} {\bibfnamefont {B.}~\bibnamefont {Pankovich}}, \bibinfo {author} {\bibfnamefont {A.}~\bibnamefont {Neville}}, \bibinfo {author} {\bibfnamefont {A.}~\bibnamefont {Kan}}, \bibinfo {author} {\bibfnamefont {S.}~\bibnamefont {Omkar}}, \bibinfo {author} {\bibfnamefont {K.~H.}\ \bibnamefont {Wan}},\ and\ \bibinfo {author} {\bibfnamefont {K.}~\bibnamefont {Br\'adler}},\ }\href {https://doi.org/10.1103/PhysRevA.110.032402} {\bibfield  {journal} {\bibinfo  {journal} {Physical Review A}\ }\textbf {\bibinfo {volume} {110}},\ \bibinfo {pages} {032402} (\bibinfo {year} {2024}{\natexlab{b}})}\BibitemShut {NoStop}%
\bibitem [{\citenamefont {Bombin}\ \emph {et~al.}(2024)\citenamefont {Bombin}, \citenamefont {Litinski}, \citenamefont {Nickerson}, \citenamefont {Pastawski},\ and\ \citenamefont {Roberts}}]{Bombin2024unifyingflavorsof}%
  \BibitemOpen
  \bibfield  {author} {\bibinfo {author} {\bibfnamefont {H.}~\bibnamefont {Bombin}}, \bibinfo {author} {\bibfnamefont {D.}~\bibnamefont {Litinski}}, \bibinfo {author} {\bibfnamefont {N.}~\bibnamefont {Nickerson}}, \bibinfo {author} {\bibfnamefont {F.}~\bibnamefont {Pastawski}},\ and\ \bibinfo {author} {\bibfnamefont {S.}~\bibnamefont {Roberts}},\ }\href {https://doi.org/10.22331/q-2024-06-18-1379} {\bibfield  {journal} {\bibinfo  {journal} {{Quantum}}\ }\textbf {\bibinfo {volume} {8}},\ \bibinfo {pages} {1379} (\bibinfo {year} {2024})}\BibitemShut {NoStop}%
\bibitem [{\citenamefont {Gottesman}(1997)}]{gottesman1997stabilizer}%
  \BibitemOpen
  \bibfield  {author} {\bibinfo {author} {\bibfnamefont {D.}~\bibnamefont {Gottesman}},\ }\href@noop {} {\emph {\bibinfo {title} {Stabilizer codes and quantum error correction}}}\ (\bibinfo  {publisher} {California Institute of Technology},\ \bibinfo {year} {1997})\BibitemShut {NoStop}%
\bibitem [{\citenamefont {Brown}\ and\ \citenamefont {Roberts}(2020)}]{PhysRevResearch2_033305}%
  \BibitemOpen
  \bibfield  {author} {\bibinfo {author} {\bibfnamefont {B.~J.}\ \bibnamefont {Brown}}\ and\ \bibinfo {author} {\bibfnamefont {S.}~\bibnamefont {Roberts}},\ }\href {https://doi.org/10.1103/PhysRevResearch.2.033305} {\bibfield  {journal} {\bibinfo  {journal} {Physical Review Research}\ }\textbf {\bibinfo {volume} {2}},\ \bibinfo {pages} {033305} (\bibinfo {year} {2020})}\BibitemShut {NoStop}%
\bibitem [{\citenamefont {Coecke}\ and\ \citenamefont {Duncan}(2011)}]{Coecke_2011}%
  \BibitemOpen
  \bibfield  {author} {\bibinfo {author} {\bibfnamefont {B.}~\bibnamefont {Coecke}}\ and\ \bibinfo {author} {\bibfnamefont {R.}~\bibnamefont {Duncan}},\ }\href {https://doi.org/10.1088/1367-2630/13/4/043016} {\bibfield  {journal} {\bibinfo  {journal} {New Journal of Physics}\ }\textbf {\bibinfo {volume} {13}},\ \bibinfo {pages} {043016} (\bibinfo {year} {2011})}\BibitemShut {NoStop}%
\bibitem [{\citenamefont {Auger}\ \emph {et~al.}(2018)\citenamefont {Auger}, \citenamefont {Anwar}, \citenamefont {Gimeno-Segovia}, \citenamefont {Stace},\ and\ \citenamefont {Browne}}]{PhysRevA.97.030301}%
  \BibitemOpen
  \bibfield  {author} {\bibinfo {author} {\bibfnamefont {J.~M.}\ \bibnamefont {Auger}}, \bibinfo {author} {\bibfnamefont {H.}~\bibnamefont {Anwar}}, \bibinfo {author} {\bibfnamefont {M.}~\bibnamefont {Gimeno-Segovia}}, \bibinfo {author} {\bibfnamefont {T.~M.}\ \bibnamefont {Stace}},\ and\ \bibinfo {author} {\bibfnamefont {D.~E.}\ \bibnamefont {Browne}},\ }\href {https://doi.org/10.1103/PhysRevA.97.030301} {\bibfield  {journal} {\bibinfo  {journal} {Physical Review A}\ }\textbf {\bibinfo {volume} {97}},\ \bibinfo {pages} {030301} (\bibinfo {year} {2018})}\BibitemShut {NoStop}%
\bibitem [{\citenamefont {Barrett}\ and\ \citenamefont {Stace}(2010)}]{PhysRevLett.105.200502}%
  \BibitemOpen
  \bibfield  {author} {\bibinfo {author} {\bibfnamefont {S.~D.}\ \bibnamefont {Barrett}}\ and\ \bibinfo {author} {\bibfnamefont {T.~M.}\ \bibnamefont {Stace}},\ }\href {https://doi.org/10.1103/PhysRevLett.105.200502} {\bibfield  {journal} {\bibinfo  {journal} {Physical Review Letters}\ }\textbf {\bibinfo {volume} {105}},\ \bibinfo {pages} {200502} (\bibinfo {year} {2010})}\BibitemShut {NoStop}%
\bibitem [{\citenamefont {Higgott}(2021)}]{higgott2021pymatchingpythonpackagedecoding}%
  \BibitemOpen
  \bibfield  {author} {\bibinfo {author} {\bibfnamefont {O.}~\bibnamefont {Higgott}},\ }\href@noop {} {\bibfield  {journal} {\bibinfo  {journal} {arXiv:2105.13082}\ } (\bibinfo {year} {2021})}\BibitemShut {NoStop}%
\bibitem [{\citenamefont {Delfosse}\ and\ \citenamefont {Nickerson}(2021)}]{Delfosse2021almostlineartime}%
  \BibitemOpen
  \bibfield  {author} {\bibinfo {author} {\bibfnamefont {N.}~\bibnamefont {Delfosse}}\ and\ \bibinfo {author} {\bibfnamefont {N.~H.}\ \bibnamefont {Nickerson}},\ }\href {https://doi.org/10.22331/q-2021-12-02-595} {\bibfield  {journal} {\bibinfo  {journal} {{Quantum}}\ }\textbf {\bibinfo {volume} {5}},\ \bibinfo {pages} {595} (\bibinfo {year} {2021})}\BibitemShut {NoStop}%
\end{thebibliography}
%

\end{document}